\documentclass[fleqn,usenatbib,useAMS]{mnras}

\usepackage{graphicx}	
\usepackage{amsmath}	
\usepackage{amssymb}	
\usepackage{multicol}        
\usepackage{bm}		
\usepackage{pdflscape}	
\usepackage{hyperref}
\usepackage[normalem]{ulem}

\DeclareMathOperator{\sech}{sech}

\usepackage[T1]{fontenc}
\usepackage{ae,aecompl}

\usepackage{newtxtext,newtxmath}

\title[LISA View of the Milky Way's bar]{The Milky Way's bar structural properties from gravitational waves}

\author[Martijn J. C. Wilhelm]{Martijn J. C. Wilhelm$^{1}$\thanks{Contact e-mail: \href{wilhelm@strw.leidenuniv.nl}{wilhelm@strw.leidenuniv.nl}}, Valeriya Korol$^{1,2}$, Elena M. Rossi$^{1}$, and Elena D'Onghia$^3$ 
\\
$^{1}$Leiden Observatory, Leiden University, PO Box 9513, 2300 RA, Leiden, the Netherlands\\
$^{2}$ Institute for Gravitational Wave Astronomy \& School of Physics and Astronomy, University of Birmingham, Birmingham, B15 2TT, UK \\ 
$^{3}$ Department of Astronomy, University of Wisconsin, 475 N Charter Street, 53704 Madison, WI, USA }
\date{Last updated 2019 September 19; in original form 2019 July 3}

\pubyear{2020}

\begin{document}
\label{firstpage}
\pagerange{\pageref{firstpage}--\pageref{lastpage}}
\maketitle

\begin{abstract}
The Laser Interferometer Space Antenna (LISA) will enable Galactic gravitational wave (GW) astronomy by individually resolving $> 10^4$ signals from double white dwarf (DWD) binaries throughout the Milky Way. Since GWs are unaffected by stellar crowding and dust extinction unlike optical observations of the Galactic plane, in this work we assess for the first time the potential of LISA to map the Galactic stellar bar and spiral arms. To achieve this goal we combine a realistic population of Galactic DWDs with a high-resolution N-body Galactic simulation in good agreement with current observations of the Milky Way. We then model GW signals from our synthetic DWD population and reconstruct the structure of the simulated Galaxy from mock LISA observations. Our results show that while the low density contrast between the background disc and the spiral arms hampers our ability to characterise the spiral structure, the stellar bar will clearly appear in the GW map of the bulge. The axis ratio derived from the synthetic observations agrees within $1\sigma$ with the reference value, although the scale lengths are underestimated. We also recover the bar viewing angle to within one degree and the bar's physical length to within 0.2 kpc. This shows that LISA can provide independent constraints on the bar's structural parameter, competitive compared to those from electromagnetic tracers. We therefore foresee that synergistic use of GWs and electromagnetic tracers will be a powerful strategy to map the Milky Way's bar and bulge.

\end{abstract}




\section{Introduction}

It is established that the Milky Way has a central stellar bar and a spiral structure that propagates through its stellar and gaseous disc. Observations of molecular masers associated with very young high-mass stars strongly suggest that the Milky Way is a four-arm spiral \citep{rei19}.
However the nature and the structure of these non-axisymmetric features remain uncertain. Indeed, the amplitude, length and pattern speed of the stellar bar are debated, and constant effort to determine their values is motivated by their importance for a broad range of Galactic studies. For example, these bar features are key to understanding the properties of the disc outside the stellar bar \citep{min10b}, the kinematics in the solar neighbourhood \citep[e.g.,][]{deh00,per17,don19,mon19}, and the observed non-circular gas flow \citep[e.g.,][]{bis03}.

Thanks to the {\it Gaia} mission \citep{gaia,dr2} -- releasing proper motions and parallaxes for almost two billion stars -- progress has been made in understanding the central region of the Galaxy, despite the strong dust extinction \citep[e.g.,][]{and19,bov19}. The current picture of the central Galactic region is that the stellar bar extends as far as $4.5-5\,$kpc from the Galactic centre and the pseudo-bulge is likely to have peanut-like shape \citep{ben05,naf10,zoc16,weg15,val16}.

Of great importance for the dynamics of the bar and its surrounding disc is the pattern speed $\Omega_{\rm P}$ of the bar itself (or equivalently its co-rotation radius), because it defines the locations of resonances associated to the bar \citep[e.g.,][table~1]{don19}. Past measurements indicated a rather fast rotation rate (pattern speed) for the Galactic bar of the order of $\Omega_{\rm P}$ = 55\,km\,s$^{-1}$\,kpc$^{-1}$ \citep{eng99,fux99,deb02,bis03}
with a bar length of $\sim$ 3\,kpc that suggested the Sun being close to the outer Lindblad resonance of the bar. 
However, recent measurements of both the three-dimensional density of Red Clump giants \citep{weg15} and the gas kinematics in the inner Galaxy \citep{sor15} point out that the bar pattern speed might be significantly slower. In particular, \citet{por17}, by modelling the kinematics and photometry of stars in the bar, inferred a pattern speed of 39\,km\,s$^{-1}$\,kpc$^{-1}$. 

Recently, \citet{san19b} derived a pattern speed of $\Omega_{\rm P} \approx 41\,$km\,s$^{-1}\,$kpc$^{-1}$, applying a modified \citet{tre84} method on a proper motion data set assembled from both multi-epoch VVV (Vista Variables in the Via Lactea) survey \citep{min10a,sai12} and {\it Gaia} data release 2 \cite[DR2,][]{dr2}. Another study derived line-of-sight integrated and distance-resolved maps of average proper motions and velocity dispersions from the VVV Infrared Astrometric Catalogue, combined with data from {\it Gaia} DR2, and found agreement with a bar pattern speed of 39 km s$^{-1}$ kpc$^{-1}$ \citep{weg15}.
A combination of {\it Gaia} DR2 and APOGEE \citep[Apache Point Observatory Galactic Evolution Experiment,][]{APOGEE} data has been also used to infer a similar bar pattern speed \citep{bov19}, although measurements may be affected by systematic uncertainties resulting from photometric distances and the small number of {\it Gaia} stars in the disc plane. The Bayesian code {\sc StarHorse}, 
combining {\it Gaia} parallax information with APOGEE spectroscopic information in addition to other photometric bands,  obtained distances in the region dominated by the bar and bulge \citep{and19}. Yet, distances are inferred with uncertainties of the order of 1\,kpc ($\sim$10 per cent precision). 

Another unresolved issue concerns the bar `viewing angle', i.e. the orientation of the Sun with respect to the long-axis of the bar. Most authors have found a viewing angle in the range between 10 -- 45$^\circ$ \citep[][figure~17]{simion_parametric_2017}. 

Recently, it has been shown that it is possible to re-construct a full 3D picture of the Milky Way by exploiting gravitational wave (GW) radiation from Galactic ultra-compact detached double white dwarf (DWD) binaries \citep{kor19}. 
Galactic DWDs can be detected in the milli-Hz GW band with the future Laser Interferometer Space Antenna \citep[LISA,][]{lisa}.
Population synthesis studies forecast $>10^4$ DWDs to be individually resolvable  by LISA \citep[e.g.][]{kor17,lam19,bre19a}. 
Such a large number of detections distributed across the Galaxy will allow us to map the Milky Way in GWs and precisely measure its structural parameters like the scale radii of the  bulge and the disc \citep{ada12,kor19}, while combined GW and optical observations of DWDs can be used to derive the mass of the bulge and the disc component of the Galaxy \citep{kor19}. 
At frequencies <3\,mHz the LISA band starts to be overpopulated by DWDs giving rise to an unresolved confusion background \citep[e.g.,][]{rob17}. 
Although affecting detectability of extra-galactic LISA sources, this background encodes the properties of the overall stellar population in the Galaxy and can also be used to recover Milky Way's parameters such as the disc scale height \citep{ben06,bre19}. 

In this work we explore for the first time the potential of GWs to characterise the Milky Way's bar's and spirals' structural properties.  To achieve this we combine a DWD binary population synthesis model with GALAKOS, a high resolution Milky Way-like galaxy simulation \citep{don19}. We consider the population of {\it detached} DWDs only, because distance determination for these binaries is not affected by systematic uncertainties due to astrophysical processes such as mass transfer. 
In Section~\ref{sec:method} we explain how we combine these two tools to obtain a realistic catalogue of Galactic DWDs. In Section~\ref{sec:mock_obs} we describe how the catalogue is processed to obtain mock LISA observations. Next, we outline the method adopted to infer the structure of the Galaxy from LISA observations (Section~\ref{sec:4}). In Section~\ref{sec:5} we present our results. Finally, we discuss how our results compare to electromagnetic observations and draw conclusions in Section~\ref{sec:6}. 

\section{Methods} \label{sec:method}

To model the population of DWDs in the Milky Way we follow the same method as in \citet{kor19}, but instead of employing an analytic gravitational potential we use high resolution numerical simulations of \citet{don19} when assigning positions to DWDs. This Galaxy model includes  a  self-consistent spiral pattern and bar whose  properties  are  in  agreement with current observations. It allows us to test how the Galactic structure will be observed with LISA.

\subsection{Binary population synthesis model} 

\begin{figure}
    \centering
    \includegraphics[width=1\linewidth]{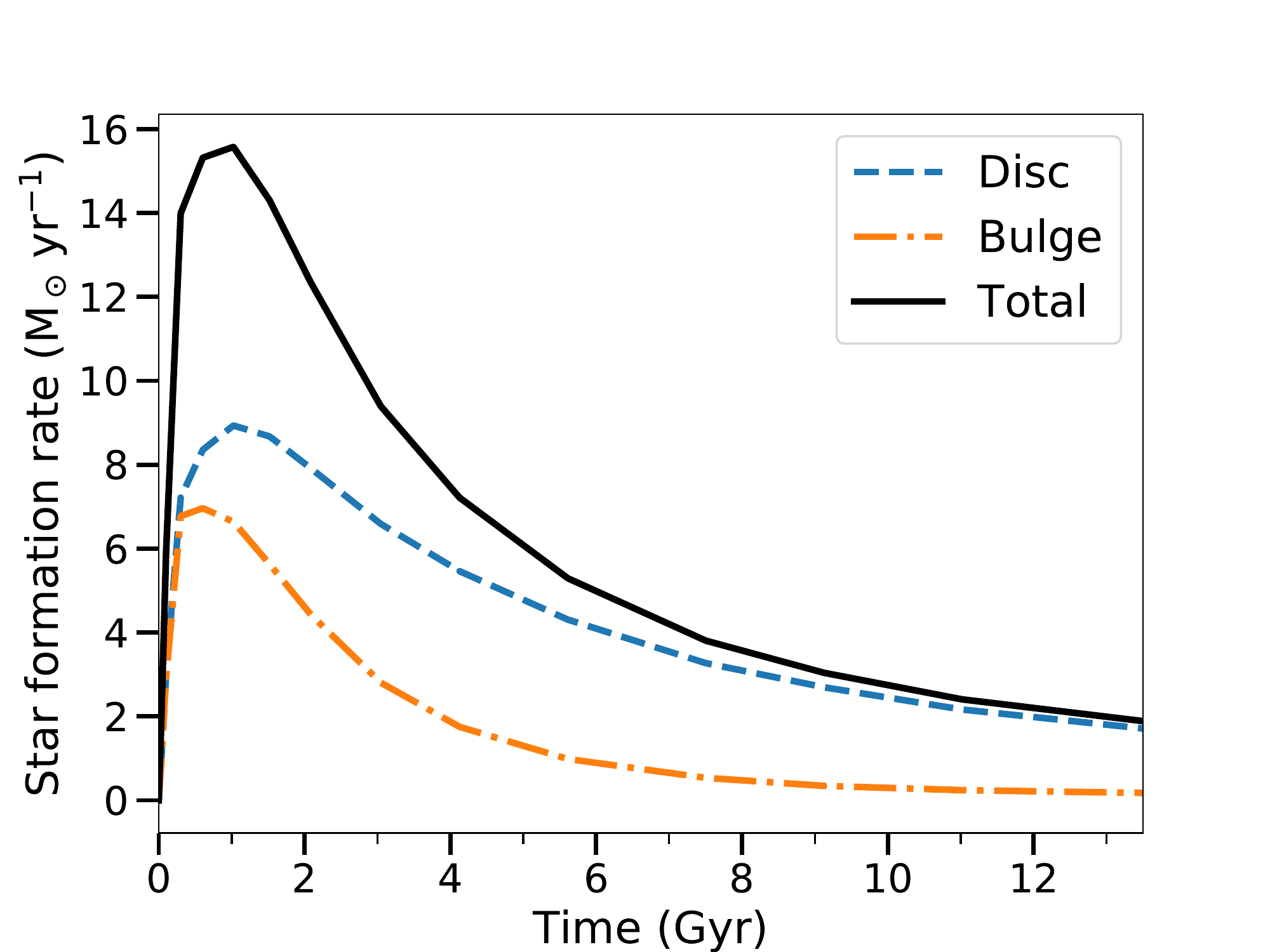}
    \caption{The star formation history giving rise to the DWD population used in this work \citep{boi99}. The present day value is 1.87\,M$_\odot$\,yr$^{-1}$.}
    \label{fig:StarFormationHistory}
\end{figure}

We employ the DWD population model from \cite{too12}, obtained using binary population synthesis code {\sc SeBa}, developed by \citet{por96}, later adapted for DWDs by \citet{nel01} and \citet{too12}. 
The progenitor population is constructed by randomly sampling binary properties with a Monte Carlo technique from distributions motivated by currently available observations for intermediate- and low-mass stars.
Specifically, the mass of the primary star is drawn from the initial mass function of \citet{Kroupa1993} in the range between 0.95 and 10$\,$M$_{\odot}$. 
The mass of the secondary star is derived from a uniform mass ratio distribution between 0 and 1 \citep{Kraus2013}. 
To draw the initial binary orbital separations and binary eccentricities we adopt respectively a log-flat and a thermal distributions \citep{Abt1983,Heggie1975,Kraus2013}. 
The initial binary fraction was set to 0.5 and the metallicity was set to the Solar value. 

First, {\sc SeBa} evolves the initial population from the Zero-age main-sequence until both stars become white dwarfs. 
Then, from DWD formation until the present time binaries are further evolved via GW emission that causes binaries' orbits to shrink \citep[e.g.,][]{pet64}. 
Finally, we remove binaries from the catalogue if they have begun mass transfer (i.e. when one of the two white dwarfs fills its Roche lobe) or they have already merged within the present time.

To assign a realistic present time age distribution of DWDs we used the star formation rate (SFR) grid from a chemo-spectrophotometric galaxy evolution model of \citet{boi99}.
Our modelling of star formation implicitly assumes that once the bar is formed, it drives gas toward the central regions of the Galaxy enhancing star formation.  This is a reasonable assumption as the  majority of the bulge forms as part of the bar, according to the recent estimates of the kinematics of the stars of the bulge and bar \citep{por17}.
Our method does not envisage a different star formation history in the Milky Ways's spiral arms. However, as the bar and spiral structure develop the stars orbiting in the disc tend to increase the amplitude of their epicycle causing an increase of the in-plane velocity dispersion and a global heating of the disc \citep{BinneyTremaine}. As a result the spiral arms that propagate through the old stellar disc formed also by the old stellar population such as DWDs will be less visible. 

For the distribution of DWDs in our Milky Way model we resampled the DWD population of \cite{kor19}, without information whether a DWD originated in the bulge or in the disc. Thus, our population has a single star formation history equal to the (weighted) sum of the disc and bulge star formation histories of \citet{boi99}, for both its disc and bulge. 

\label{sec:MW model}
\begin{figure*}
    \centering
    \includegraphics[width=0.8\linewidth]{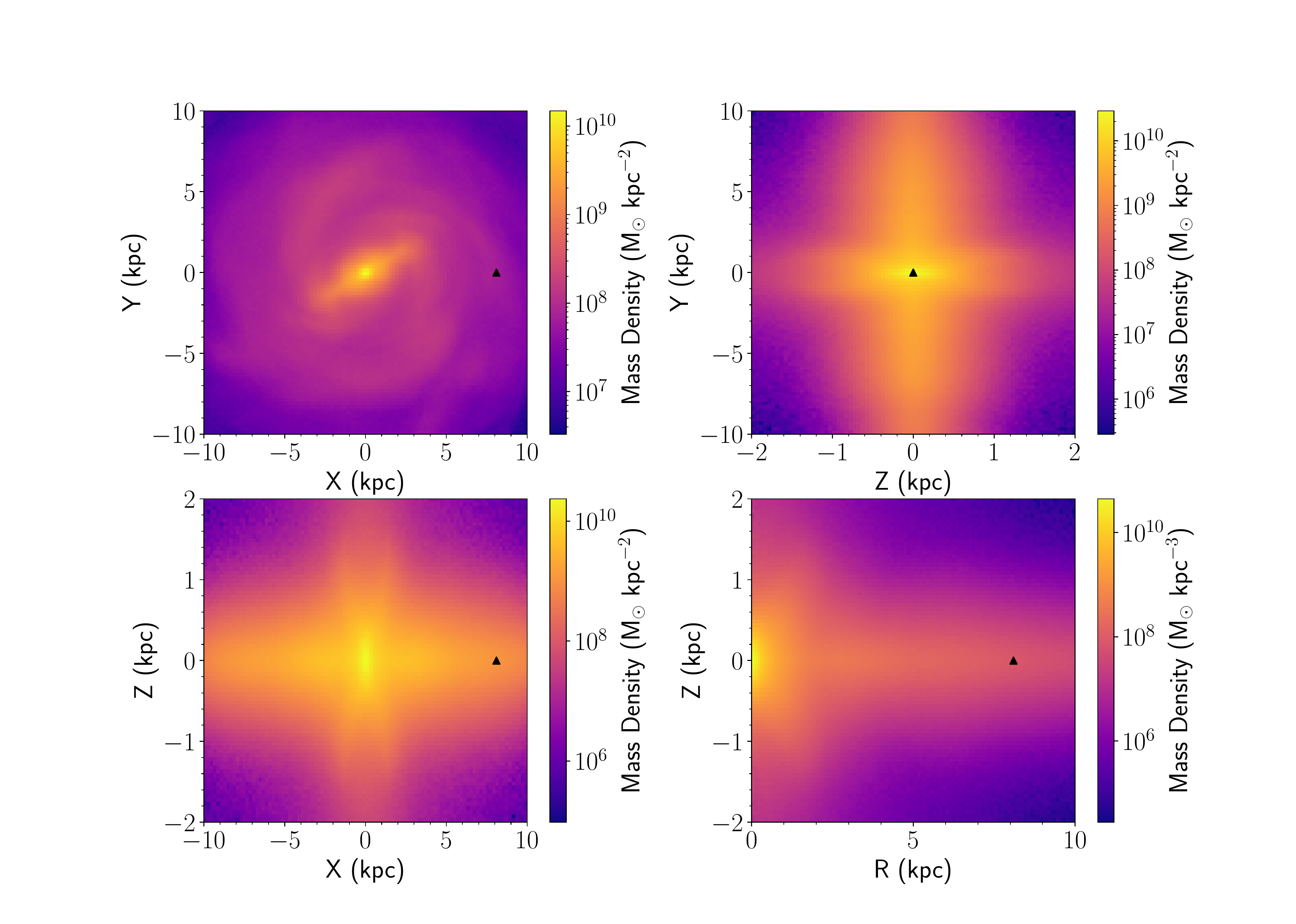}
    \caption{Projected mass density maps of the Galaxy model, in the $X-Y$ plane, $Z-Y$ plane, $X-Z$ plane, and $R-Z$ plane, where $R^2 = X^2 + Y^2$ and $(X, Y, Z)$ is the triad of Cartesian coordinates centred on the Galactic centre. Included are the bulge and disc components. The black triangle marks the Sun's position. Our line of sight makes an angle of $\sim 30$ degrees with respect to the bar's long axis. }
    \label{fig:galaxy_maps}
\end{figure*}

One of the most impacting assumption in binary population synthesis is the prescription for the common envelope (CE) evolution.
CE is a short phase of the binary evolution ($\sim10^3\,$yr) in which the more massive star of the pair expands and engulfs its companion \citep{Paczynski1976,Webbink1984}. 
During this phase the binary orbital energy and angular momentum can be transferred to the envelope, due to the dynamical friction experienced by the companion star when moving through the envelope. This process continues until either the binary merges or the envelope is ejected from the system leaving behind the core of the expanded star and its companion in a tighter orbit. Typically, the CE is implemented in the binary population synthesis either by parametrising the conservation equation for energy (through the $\alpha$ parameter) or the balance equation of the system's angular momentum (through the $\gamma$ parameter) \citep[see][for a review]{Ivanova2013}.
For this study we adopt the $\gamma \alpha$ DWD evolution model, in which both parametrisations are allowed: the $\gamma$-prescription is applied unless the binary contains a compact object or the CE is triggered by a tidal instability, in which case $\alpha$-prescription is used.
Synthetic catalogues of DWDs produced using this evolution model have been carefully calibrated against state-of-the-art observations of DWDs in terms of both mass ratio distribution \citep{too12} and number density \citep{too17}.
Future optical surveys such as the Vera Rubin Observatory \citep{ive08} will provided large samples of new short-period DWDs that will help to further constrain DWD evolution models \citep{kor17}.

\subsection{Milky Way model}

To simulate the Milky Way we use a snapshot of {\sc GALAKOS}, a high-resolution N-body simulation of a stellar disc with structural parameters that reproduce the currently observed  properties  of our Galaxy \citep{don19}. The simulation was carried out with {\sc GADGET3}, a parallel TreePM-Smoothed particle hydrodynamics (SPH) code developed to compute the evolution of stars and dark matter, treated as collisionless fluids. The phase space is discretised into fluid elements that are computationally realised as particles in the simulation. The total number of N-Body particles employed in the simulation is 90 million. The gravitational softening length adopted is 40 pc for the dark halo, 28 pc for the stellar disc and 80 pc for the bulge. Note that the gas component was not included in the simulation.

The Milky Way model consists of three components: a dark matter halo, a rotationally supported stellar disc, and a spherical stellar bulge. These components are modelled as follows:
\begin{itemize}
    \item {\it Dark matter halo} is modelled with the \citet{her90} density profile
    \begin{equation}
    \rho_{\rm DM}(r) = \frac{M_{\rm DM}}{2\pi}\frac{a}{r(r+a)^3}
\end{equation}
where $M_{\rm DM}=1 \times 10^{12}\,$M$_{\odot}$ is the total dark matter mass and $a=30\,$kpc is the radial scale length. This choice is motivated by the fact that in its inner part, the shape of the
density profile is identical to the Navarro-Frenk-White fitting formula of the mass density distribution of dark matter halos inferred in cosmological simulations \citep{nav96}. The dark matter mass in 
{\sc GALAKOS} is sampled with 60 million particles.

\item {\it Stellar disc} is represented by an exponential radial stellar disc profile with an isothermal vertical distribution
\begin{equation}  \label{eqn:disc}
\rho_{\rm disc}(R,z) = \frac{M_{\rm d}}{8 \pi z_0 R^2_{\rm d}} e^{-R/R_{\rm d}}\sech^2 ( z/z_0), 
\end{equation} 
where $M_{\rm d}=4.8 \times 10^{10}\,$M$_{\odot}$ is  the  disc  mass,  $R_{\rm d}=2.67\,$kpc is  the  disc  scale  length,  and  $z_0=320\,$pc is  the  disc  scale  height.  The disc mass is discretised with 24 million particles. The gas component is not included in the model.

\item {\it Stellar bulge} is also described by the Hernquist model. 
The total mass of the bulge adopted in the simulation is $M_{\rm b}=8 \times 10^9$M$_{\odot}$ with a scale length $a_{\rm b}=320\,$pc. The number of particles sampled in the bulge is 8.4 million. Note, that the bulge in this simulation does not rotate.
\end{itemize}

The simulation results in a Milky Way-like galaxy with the total stellar mass of $5.6 \times 10^{10}\,$M$_{\odot}$ and  accounts  for  a  time-varying  potential that after $2.5\,$Gyrs forms a bar with a length of $4.5\,$kpc, a width of $\approx 2.5$\,kpc and a pattern speed of $40\,$km\,s$^{-1}\,$kpc$^{-1}$ (see Appendix in \citet{don19} for the details). Figure~\ref{fig:galaxy_maps} shows the Milky Way  model under different projections. Note that this model includes two prominent spiral arms, whereas the Milky Way is believed to have four. 

To calculate the DWD density distribution of the Galaxy (which we assume to be the stellar number density distribution) we use the kernel density estimation (KDE) method. 
We approximate the mass-density distribution of the Galaxy by placing a function called a {\it kernel}, in our case a three dimensional Gaussian, at the position of every simulation particle. The superposition of these functions is then used as a probability density function to populate the simulated Galaxy with DWD binaries. In other words, this is equivalent to assigning a DWD's position around a random simulation particle with a Gaussian probability density centred on the particle. We use a bandwidth of the KDE (equivalently the standard deviation of the Gaussian) of $10\,$pc.
This choice minimises the residuals between the original simulation particle distribution and that reconstructed (backwards) from the KDE matter density distribution. In addition, this bandwidth allows us realistically describe the white dwarf's spatial distribution, whose local space density is $0.0045\,$pc$^{-3}$ \citep[e.g.,][]{hol18}.

\subsection{GW emission from simulated binaries}

Binary population synthesis provide us orbital periods $P_{\rm orb}$ and component's masses $m_1$ and $m_2$, while the Galaxy simulation gives us the DWDs' 3D positions. Here we provide relations between these quantities and those characterising the GW signal of DWD binaries: GW frequency $f$, its time derivative $\dot{f}$, the GW amplitude ${\cal A}$, the source latitude and longitude in the sky, as seen by LISA $(\bar{\theta}, \bar{\phi})$, binary orbital plane inclination with respect to the line-of-sight $\iota$ , the angle $\psi$ between the two wave polarisations, and finally the initial orbital phase $\phi'_0$.
The first three are obtained from binary properties as
\begin{equation}
    f= \frac{2}{P_{\rm orb}},
\end{equation}
\begin{equation}
    \dot{f} = \frac{96}{5}\pi^{8/3} \left( \frac{{\cal M}G}{c^3}\right)^{5/3}f^{11/3}
\end{equation} and
\begin{equation}
{\cal A} = \frac{2 (G {\cal M})^{5/3}(\pi f)^{2/3}}{c^4d},
\end{equation}
where ${\cal M} = (m_1 m_2)^{3/5}/(m_1 + m_2)^{1/5}$ is the chirp mass and $d$ is the luminosity distance, $G$ and $c$ are respectively the gravitational constant and the speed of light. We compute sky coordinates and the distance to the source in the LISA reference frame as
\begin{equation}
\begin{aligned}
\bar{\theta} &= {\pi}/{2} - \arccos({z_{\rm ecl}}/{d})  \\
\bar{\phi} &= \arctan({y_{\rm ecl}}/{x_{\rm ecl}})
\end{aligned}
\end{equation}
with $(x_{\rm ecl}, y_{\rm ecl}, z_{\rm ecl})$ being the triad of Cartesian coordinates in the heliocentric ecliptic system and $d = \sqrt{x_{\rm ecl}^2+y_{\rm ecl}^2+z_{\rm ecl}^2}$. 

We assume the distance of the Sun from the Galactic centre to be of $8.1\,$kpc \citep{abu19,rei19} and a viewing angle with respect to the bar's long-axis of $30^\circ$. Finally, binary inclination angles are drawn from a probability density function uniform in $\cos \iota$ so to ensure that all viewing angles between the orbital plane and our line of sight are equally probable. Since there is no preferential polarisation angle and initial orbital phase, $\psi$ and $\phi'_0$ are randomly drawn from a uniform distribution between $[0, \pi]$ and $[0, 2\pi]$ respectively.


\section{Mock GW observations} \label{sec:mock_obs}

\begin{figure}
    \centering
    \includegraphics[width=1\linewidth]{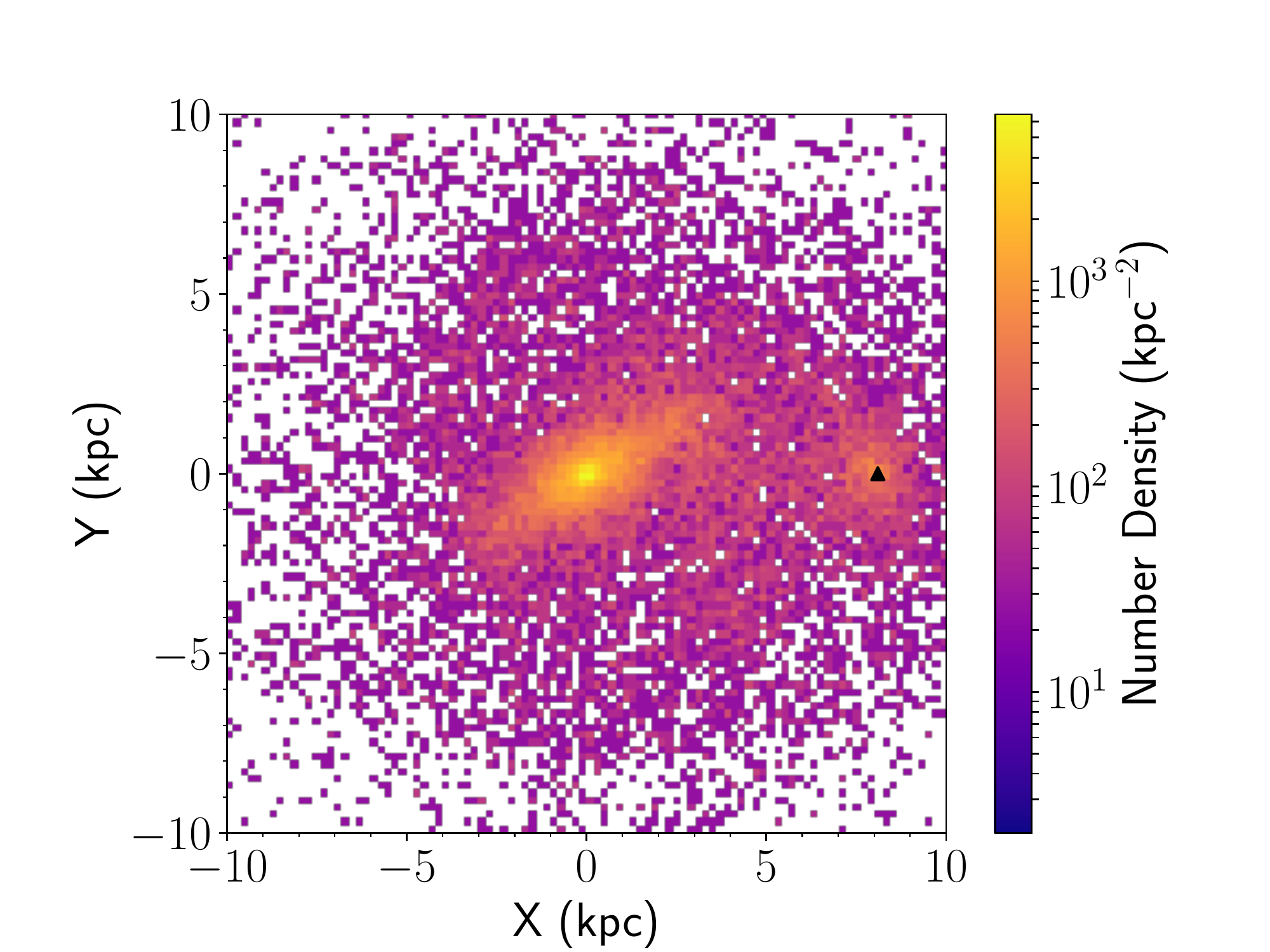}
   \includegraphics[width=1\linewidth]{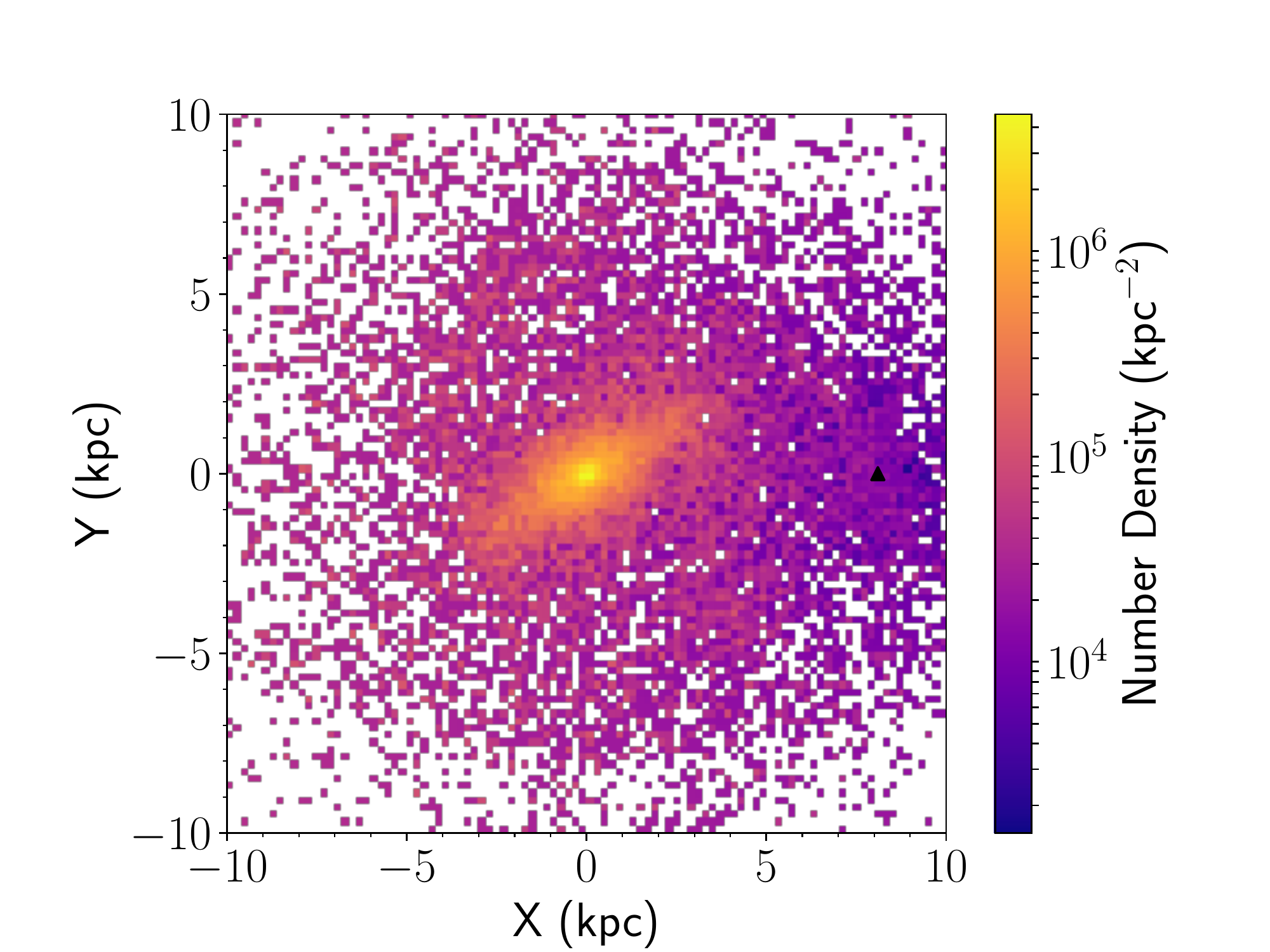}
    \caption{Projected number density maps of DWDs individually detected with LISA in $4\,$yr of mission projected on the Galactic plane ({\it top panel}) and the same map corrected for the distance bias using Eq.~\eqref{eq:bias} ({\it bottom panel}). The black triangle indicates the location of the LISA detector at $X=8.1$\,kpc, and $Y=0$.}
    \label{fig:GalMaps1}
\end{figure}

In this work we consider the latest LISA mission concept consisting of three identical spacecrafts in an equilateral triangle configuration of $2.5 \times 10^6\,$km per side \citep{lisa}. 
The spacecrafts are designed to exchange laser links (2 links per arm of the triangle) closing the triangular configurations. 
This design allows generation of two sets of data streams yielding two independent time-series with uncorrelated noise, in this way maximising the signal-to-noise ratio (SNR) of a GW event \citep{val05}.
We assume the nominal detection threshold corresponding to a signal-to-noise ratio (SNR) of 7, a mission duration of $4\,$ years and instrument noise model from the LISA mission proposal \citep{lisa}. 

To compute the number of LISA detections and predict the uncertainties on their parameters we employ the Mock LISA Data Challenge (MLDC) pipeline developed by \citet{lit13}\footnote{The pipeline is publicly available at \url{github.com/tlittenberg/ldasoft}.} and more recently updated by A. Petiteau.
The pipeline is designed for optimising the analysis of a large number of quasi-monochromatic GW sources simultaneously present in the data. To extract LISA detections from the input catalogues the pipeline iteratively computes the overall noise level by running a median smoothing function through the power spectrum of the population and extracts high SNR sources until convergence. Finally, the pipeline computes the errors on GW observables by computing Fisher information matrices for detected sources. 

We find $21.7 \times 10^3$ DWDs with SNR $>7$, i.e. individually resolved by LISA, in agreement with our previous results \citep{kor19}. 
We show their number density maps projected in the Galactic plane in Fig.~\ref{fig:GalMaps1} (upper panel); we indicate the position of the LISA detector with the black triangle. 

\subsection{Distance bias} \label{sec:bias}

 In contrast to electromagnetic studies, GW observations provide direct measurements of the amplitude of the waves, rather than the energy flux. This implies that the observed signal scales  as $1/d$, rather than $1/d^2$, allowing to generally detect sources in GWs at larger distances than in the more traditional observational bands. Still, we will progressively lose more and more distant sources, specifically those emitting at frequencies less than $3$ mHz \citep{lam19,kor20,roe20}. In order to reconstruct the unbiased spatial distribution from the observed one, we therefore have to correct for the missing binaries, and we do so by following the approach outlined in \citet[][eqs.~(B1)-(B3)]{kor19}. We briefly describe this method below.

We assume that the population of DWDs is homogeneous in its binary properties throughout the Galaxy and that the population evolves in frequency only due to GW emission, thus ignoring the effect of star formation on the DWD frequency distribution. Under these assumptions the fraction of observed sources with respect to the total (input catalogue) becomes a function of distance only. 
 First, we compute the detection fraction in $\sim250$ spherical shells around the Sun spaced logarithmically between 0.1 and 100\,kpc; we count the number of detected sources in each shell and divide it by the total number of sources present there. We then fit the resulting distribution with a power law in distance (that we call the de-biasing function). Then, the inverse of this power law is used to assign a weight to each detected source as
\begin{equation} \label{eq:bias}
    w\left(d\right) \approx 1.01\times 10^2 \left(\frac{d}{1\text{\,kpc}}\right)^{0.94}.
\end{equation}
Effectively, each weight $w(d)$ represents the number of non-detections for each detected source at a distance $d$. 
The effect of the correction is clearly visible in the bottom panel of Fig.~\ref{fig:GalMaps1}, where there are relatively fewer sources around the Sun with respect to the upper panel.

The two maps in Fig.~\ref{fig:GalMaps1} reveal the potential of using GW sources as tracers of Galactic structure that allow the reconstruction of the whole surface of the disc including the far side of the Milky Way. 
In particular, the denser central Galactic regions are well mapped and the shape of the bar is already clearly visible without further data processing, unlike for optical observations \citep[e.g.,][]{and19}. Only hints of the spiral structure are present at distances $< 5\,$kpc from the Galactic Centre, but overall it cannot be clearly identified from visual inspection alone.


\section{Galaxy structure analysis} \label{sec:4}

In this section we outline the method adopted to recover the properties of our fiducial Galaxy model from mock LISA observations. We carry out the analysis of the density distribution in Fourier space. First, we give the definition of the Fourier modes and discuss their meaning in our study. Next, we consider simple analytic distributions useful for understanding results presented in the next Section.

To define Fourier modes of the Milky Way's density distribution, we integrate over concentric rings centred on the Galactic Centre. Within these rings, both the bar and spiral arms are expected to appear as periodic over-densities. We define the Fourier modes $A_m$ as

\begin{equation} \label{eq:fourier}
    A_m\left(R\right) = \int_0^{2\pi} n\left( R,\phi \right)\exp\left( im\phi \right)d\phi,
\end{equation}
where $m$ denotes the angular frequency of the mode, $R$ is the cylindrical Galactocentric radius, $\phi$ is the polar angle in the Galactic plane, and $n\left( R,\phi \right)$ is the number density distribution projected on the Galactic plane. 
A prominent $A_2$ mode implies a pair of over-densities and thus implies either a bar or a pair of spiral arms. 
The difference between the bar and spiral arm structures is encoded in the mode (complex) phase. 
A bar has a complex phase which is  constant with $R$, while spiral arms 
are characterised by a radial dependent phase. 
In the following we normalise the Fourier modes $A_2$ to the zero mode $A_0$, representing the total mass in the ring, so that the ratio is confined between  $\left[0,1\right]$.

Next, we analytically derive the Fourier modes for a number of simple analytic density distributions: a uniform density bar, logarithmic spirals, a toy barred spiral galaxy and a 2D Gaussian density bar.
These models are useful for interpreting the results when analysing the density distribution of the DWDs detected by LISA.

\subsection{Uniform density bar} \label{sec:unifom_bar}


\begin{figure}
    \centering
    \includegraphics[width=0.5\linewidth]{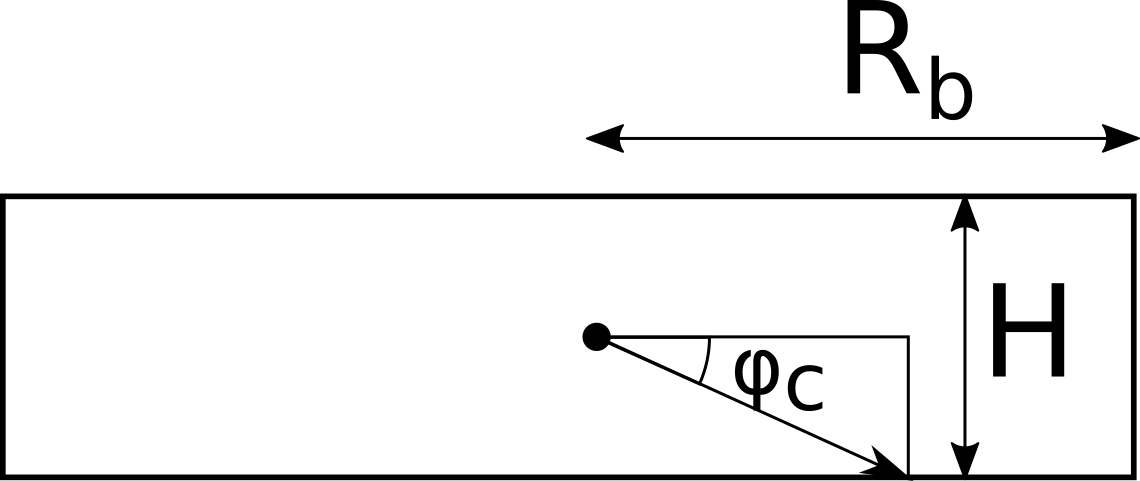}
    \caption{A sketch of a uniform bar with half-length $R_{\rm b}$ and full width $H$. From geometrical considerations we can see that $\phi_{\rm c}\left(R\right) = \arcsin\left(\frac{H}{2R}\right)$}
    \label{fig:UniformBar}
\end{figure}

\begin{figure*}
    \includegraphics[width=0.8\linewidth]{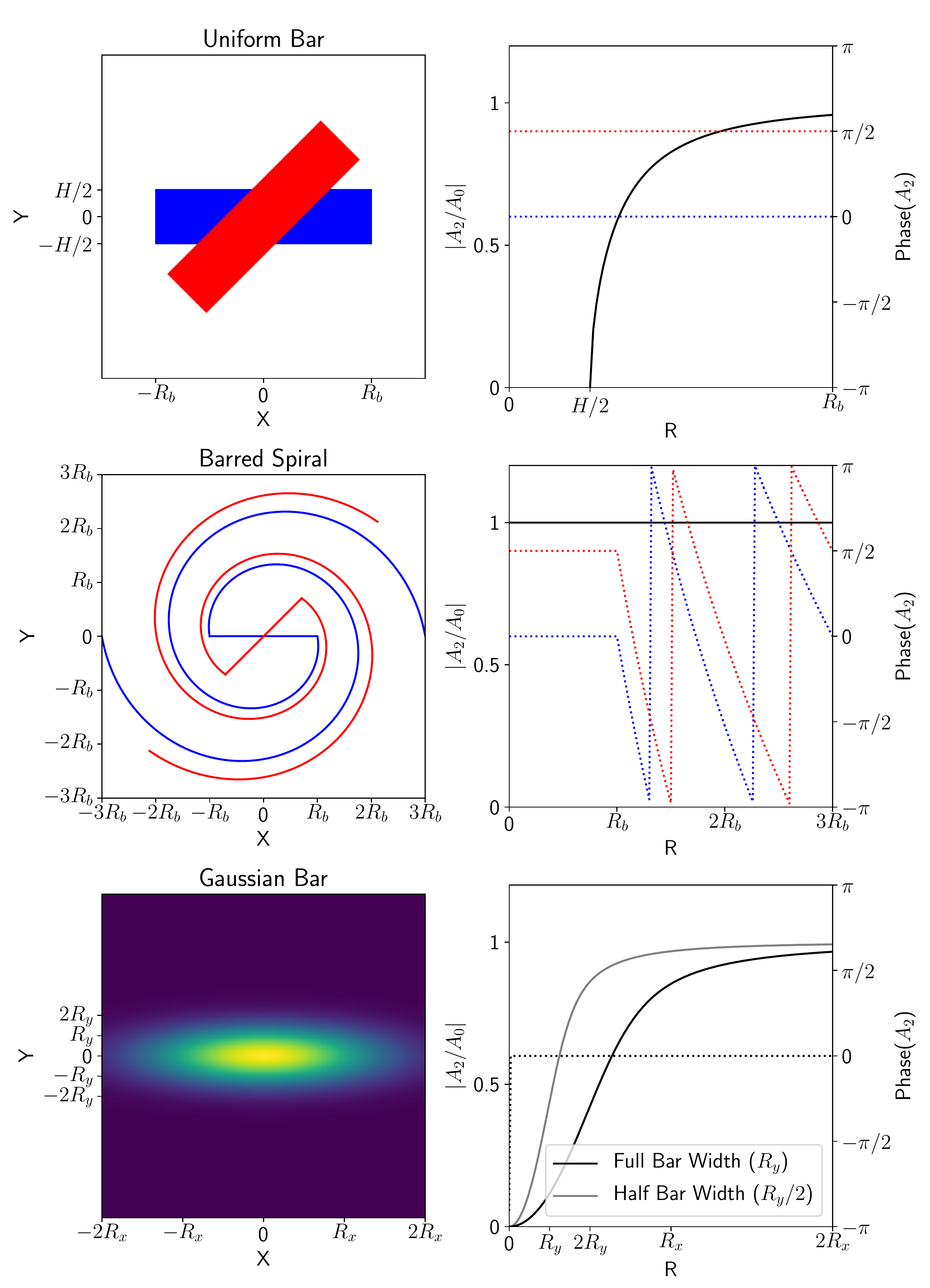}
    \caption{ Three examples of density distributions (left panels) with (in the right panels) the respective plots of the normalised $m=2$ Fourier mode magnitude (solid lines) and its complex phase (dotted lines) as a function of $R$. {\bf Top panels}: {\it A uniform density bar} under two different rotations. The magnitude vanishes for radii smaller than half the bar width, and then quickly converges to 1. The phase is constant for all radii, and is shifted by twice the angle to the $x$-axis. {\bf Middle panels}: {\it A toy barred spiral} with infinitesimally thin bar and logarithmic spirals. In this case the magnitude is constant for all radii and equal to 1, while the phase is constant for radii smaller than the bar length and drops beyond. The jumps in the phase are due to phase-wrapping to the interval  $\left(-\pi,\pi\right]$. {\bf Bottom panels}: {\it A 2D Gaussian bar density model}. The grey solid line denotes the profile of a bar with half the semi-minor scale length of the model shown. The magnitude is 0 only at the origin and converges to 1. The speed of this convergence depends on the bar semi-minor scale length. The phase is constant at 0 (except at the origin where it is undefined) like the uniform bar. }
    \label{fig:FourierIllustrations}
\end{figure*} 


Let us consider a bar with a uniform density $n_0$, half-length $R_{\rm b}$ and full width $H$. The density profile can be expressed as
\begin{equation}
    n\left(X,Y\right) = n_0 \begin{cases}
        1\quad |X|<R_{\rm b},\quad |Y| < H/2 \\
        0\quad\text{otherwise.}
    \end{cases}
\end{equation}
For certain radii $R$ there is a critical angle $\phi_{\rm c}(R) = \arcsin (H/2R)$ that indicates the boundaries of the bar in polar coordinates. This is represented in Fig.~\ref{fig:UniformBar}. 
Using the definition in Eq.~\eqref{eq:fourier} we can write the Fourier modes as
\begin{equation}
\begin{split}
    A_m\left(R\right) & = n_0\left(\int_{-\phi_c}^{\phi_c} \exp\left(im\phi \right)d\phi + \int_{-\phi_c+\pi}^{\phi_c+\pi} \exp\left(im\phi\right)d\phi\right) \\
    & = n_0 \left( 1 + e^{im\pi} \right) \sin\left( m\arcsin\left( \frac{H}{2R} \right) \right).
\end{split}
\end{equation}
This only holds for $m>0$, and the absolute value of the radial coordinate $H/2 < |R| < R_{\rm b}$. Inside $H/2$ and far outside $R_{\rm b}$ the Fourier modes vanish.
Note that there is an intermediate regime between $R = R_b$, where the circle with radius $R$ covers the bar from $-\phi_c$ to $\phi_c$, and $R = \sqrt{R_{\rm b}^2 + \left(H/2\right)^2}$, where the bar is entirely inside the circle (and the Fourier modes vanish). We note that in this regime the integration boundaries will have a more complex expression; although a detailed description is beyond the scope of the paper. 

The $m=0$ mode is
\begin{equation}
    A_0\left(R\right) = 4n_0\arcsin\left(\frac{H}{2R}\right),
\end{equation}
so the normalised Fourier $m=2$ mode is
\begin{equation}
    \frac{A_2}{A_0} = \frac{1+e^{i2\pi}}{2}\frac{\sin\left(2\phi_{\rm c}\right)}{2\phi_{\rm c}}.
\end{equation}
Rotation of the bar by an angle $\phi_0$ implies a shift in the integration boundaries by the same angle. This results in a constant (complex) phase factor for the Fourier modes, i.e.
\begin{equation}
    A_2\left(R,\phi_0\right) = e^{i2\phi_0}A_2\left(R,0\right).
\end{equation}
See the top left panel of Fig.~\ref{fig:FourierIllustrations} to visualise the density profile of a uniform density bar with $\phi_0=0$ in blue and $\phi=\pi/4$ in red. 
The top right panel of Fig.~\ref{fig:FourierIllustrations} shows the normalised $m=2$ Fourier mode magnitude (solid black line) and the phase (dotted lines) for two configurations.
For $R<H/2$ the magnitude is zero and for $R>H/2$ it quickly rises and converges to 1 at $R=R_{\rm b}$. In contrast, the phase of the mode remains constant at all radii with a value equal to $2\phi_0$ as discussed above.

\subsection{Logarithmic spirals}

Although both simulations and observations do not exactly follow
logarithmic spiral structure, here we consider this example as a helpful analytic tool to construct a simple barred spiral model in Section~\ref{sec:barred_spiral}.

The logarithmic spiral can be parametrised as
\begin{equation}
    R\left(\phi\right) = R_0 \, e^{b\phi},
\end{equation}
where $R_0$ is the distance to the origin for $\phi=0$ and fixes the phase of the spiral, and $b$ is related to the pitch angle. The pitch angle is defined as $\mu = \arctan|\frac{1}{R}\frac{dR}{d\phi}|$, thus it can be expressed in terms of $b$ as $\mu=\arctan|b|$.  
The density distribution of a set of $N$ evenly distributed, infinitesimally thin logarithmic spirals is then 
\begin{equation}
    n\left(R,\phi\right) = \sum_{n=0}^{N-1}M\delta\left[ R - R_0\exp\left( b\left( \phi - 2\pi n/N \right) \right) \right].
\end{equation}
Again, using Eq.~\eqref{eq:fourier} we can express the $m=2$ mode of this density distribution as
\begin{equation}
\begin{split}
    A_2\left(R\right) & = M\sum_{n=0}^{1}\int_0^{2\pi} \delta\left[ R - R_0\exp\left(b\left( \phi - \pi n \right)\right) \right] e^{i2\phi}d\phi \\
    & = 2M\exp\left( \frac{2i}{b}\ln\frac{R}{R_0} \right)
\end{split}
\end{equation}
From this equation it follows that the normalised $m=2$ Fourier mode is 
\begin{equation}
    \frac{A_2}{A_0} = \exp\left( \frac{2i}{b} \ln\frac{R}{R_0} \right).
\end{equation}

\subsection{Barred spiral} \label{sec:barred_spiral}

Now we combine the uniform density bar and the logarithmic spirals to construct a toy model for a barred spiral galaxy. 
We consider the bar to be infinitesimally thin (i.e. $\phi_{\rm c}\rightarrow 0$) out to some radius $R_{\rm b}$ and a pair of logarithmic spirals starting from the edges of the bar (i.e. we set $R_0=R_{\rm b}$). 
The magnitude of the normalised $m=2$ Fourier mode is 1 and its phase $\Phi$ is of the form
\begin{equation}\label{eq:barred_spiral}
    \Phi\left[ \frac{A_2}{A_0}\left(R\right) \right] = \begin{cases}
    0\quad R < R_{\rm b}, \\
    \frac{2}{b}\ln\frac{R}{R_{\rm b}}\quad R > R_{\rm b}.
    \end{cases}
\end{equation}
From which follows that rotating this toy galaxy by an angle of $\phi_0$ results in a phase shift of $2\phi_0$ (considering the bar example and continuity).
This is represented in the middle panels of Fig.~\ref{fig:FourierIllustrations}. Note that the spiral arms rotate clockwise here, analogous to our galaxy model, which introduces a minus sign in Eq.~\ref{eq:barred_spiral}.
The middle right panel shows that the magnitude of the normalised $m=2$ mode (solid line) is equal to 1 at all radii, while the phase (dotted lines) is constant for $R<R_{\rm b}$ and drops afterwards. The discontinuities in the phase are due to its wrapping to the interval $\left(-\pi,\pi\right]$.

\subsection{Gaussian bar} \label{sec:gauss_bar_model}

Finally, we consider a more realistic example of a bar with a 2D Gaussian density distribution:

\begin{equation}
\begin{split}
    n\left(R,\phi\right) & = n_0 \exp \left( -\frac{1}{2}\left( \left(\frac{x\left(R,\phi\right)}{R_{\rm x}}\right)^2 + \left(\frac{y\left(R,\phi\right)}{R_{\rm y}}\right)^2 \right) \right) \\
    & = n_0 \exp \left( -\frac{1}{2R_{\rm x}^2}\left( \left(R\cos\phi\right)^2 + \left(\frac{R_{\rm x}}{R_{\rm y}}\right)^2 \left(R\sin\phi\right)^2 \right) \right).
\end{split}
\end{equation}
The Fourier modes for a Gaussian bar can then be written as
\begin{equation} \label{sec:gauss_bar}
    A_2 (R) = \int_{0}^{2\pi} n_0 \exp \left( -\frac{1}{2R^2_{\rm x}}\left( \left(R\cos\phi\right)^2 + k\left(R\sin\phi\right)^2\right) + i2\phi \right) d\phi,
\end{equation}
where $k=(R_{\rm x}/R_{\rm y})^2$ with $R_{\rm x}$ and $R_{\rm y}$ being characteristic scales in the $X$- and $Y$-directions of the bar's reference system.

Useful for later, 
we define the bar semi-major scale length $R_+$ and the bar semi-minor scale length $R_-$ as the longer and shorter scale lengths:
\begin{equation} \label{eq:rb}
    R_{+} = \text{max}\left(R_{\rm x}, R_{\rm y}\right) = \begin{cases}
        R_{\rm x}\quad\quad\quad ~\text{if}~ k>1,\quad \\
        R_{\rm x}/\sqrt{k}\quad ~\text{ if}~k<1,
    \end{cases}
\end{equation}
\begin{equation} \label{eq:h}
    R_- = \text{min}\left(R_{\rm x}, R_{\rm y}\right) = \begin{cases} 
        R_{\rm x}/\sqrt{k}\quad ~\text{ if}~ k>1,\quad \\
        R_{\rm x} \quad\quad\quad ~\text{if}~k<1.
    \end{cases}
\end{equation}

Note that a bar rotated by 90$^\circ$ has $R_{\rm x}$ and $R_{\rm y}$ flipped, but it will have the same Fourier magnitude profile as the rotation is encoded in the phase. From Eq.~ \eqref{eq:rb}-\eqref{eq:h} it follows that the axis ratio $R_-/R_{+}$ is $1/\sqrt{k}$ for $k > 1$, and $\sqrt{k}$ if $k<1$.

The integral in Eq.~\eqref{sec:gauss_bar} does not have an analytic solution and needs to be integrated numerically. We plot the numerical results in the lower panels of Fig.~\ref{fig:FourierIllustrations}.
The grey and the black solid lines in the right panel represent the trend of $|A_2/A_0|$ with $R$ respectively for the case of half bar semi-minor scale length and full bar semi-minor scale length. Both start from 0 in the origin and converge to 1 at $\geq 2 \times$ characteristic scale radius ($R_{\rm x}$ and $R_{\rm y}$). It is evident from the two examples that the amplitude of the thinner bar approaches unity faster.
The phase (dotted line) is constant at 0 (except at the origin) as in the case of the uniform bar (top panels in Fig.~\ref{fig:FourierIllustrations}). 

\subsection{Milky Way simulation}\label{fourier mw}

\begin{figure}
    \centering
    \includegraphics[width=1\linewidth]{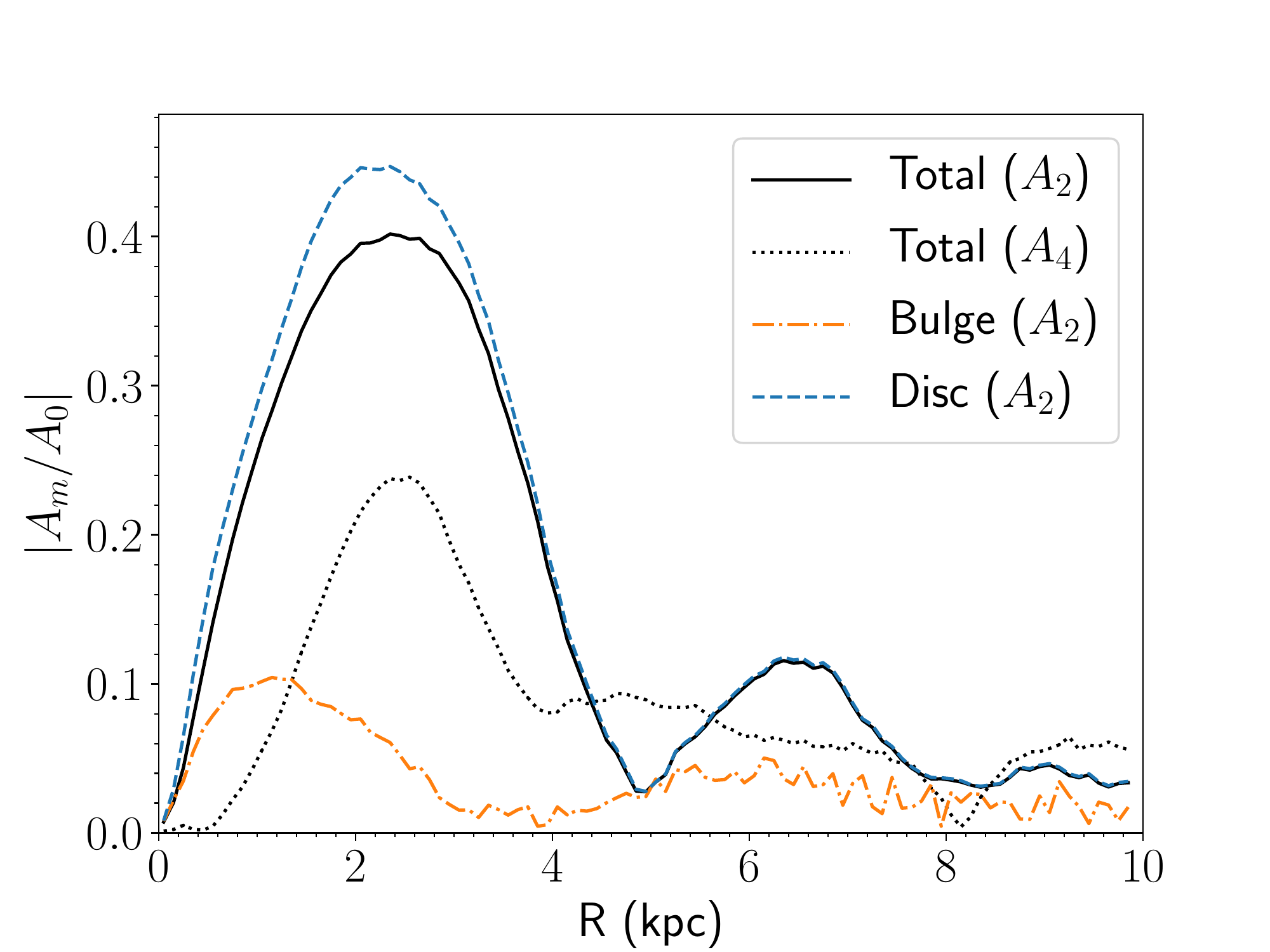}
    \includegraphics[width=1\linewidth]{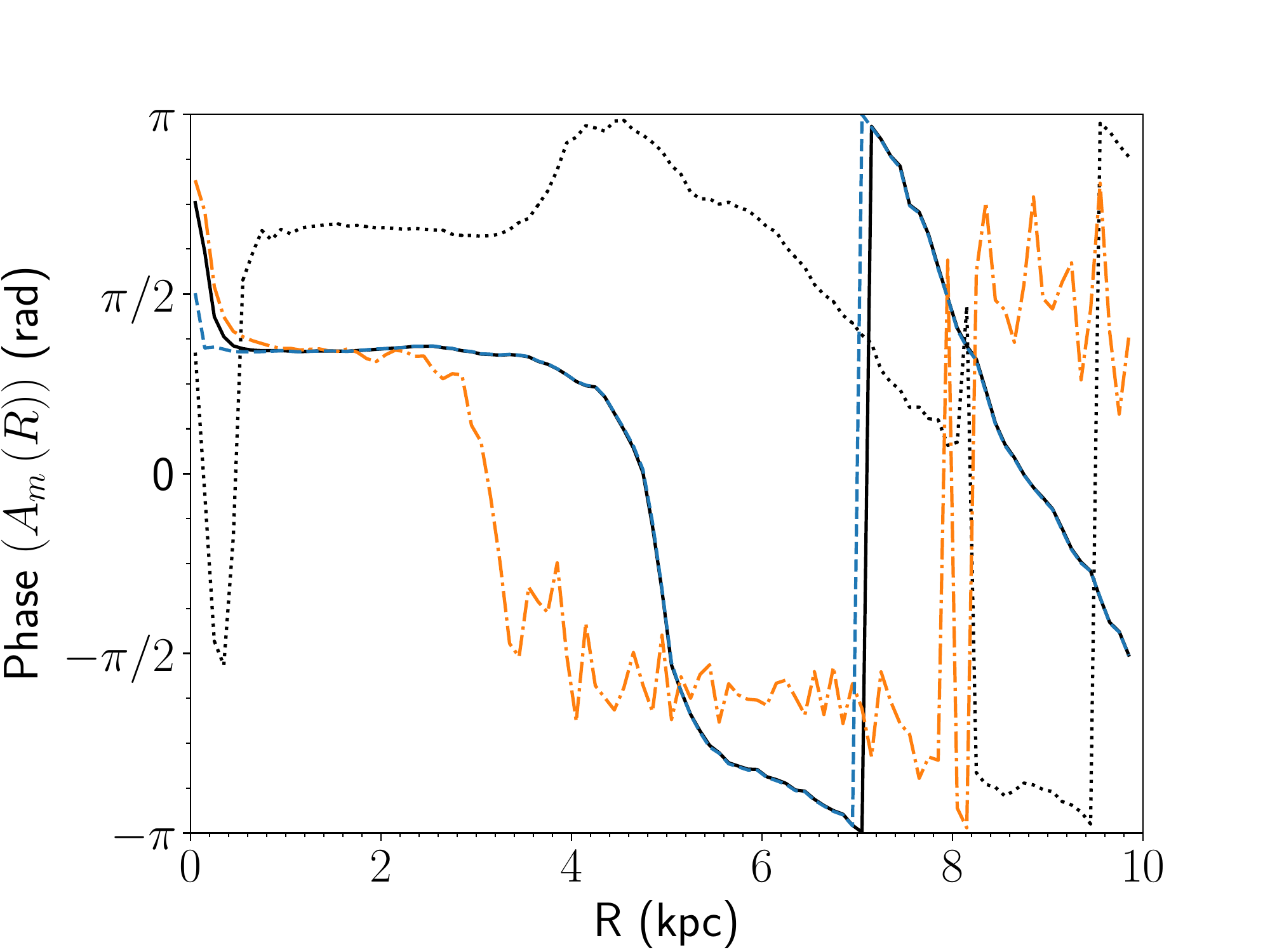}
    \caption{The Fourier magnitude (top) and phase (bottom) profiles ($m=2$) for the full galaxy model (Total $A_2$), the particles in the disc component (Disc), and the particles in the bulge component (Bulge), and for the $m=4$ mode for the full galaxy model (Total $A_4$). }
    \label{fig:fourier_galaxy_components}
\end{figure}

To compute the density profile of the simulated Galaxy (Fig.~\ref{fig:galaxy_maps}) we apply the formal definition of the density distribution for a collection of point particles given by a superposition of delta functions 
\begin{equation}
    n\left(R,\phi\right) =\sum_i \delta\left(R - R_i\right)\delta\left(\phi - \phi_i\right),
\end{equation}
where the sum is performed over all detections $i$.
Entering this in Eq.~\eqref{eq:fourier} we recover the expression for the Fourier modes: 
\begin{equation}
\begin{split}
    A_m\left(R\right) & = \int_{0}^{2\pi} \sum_i \delta\left(R - R_i\right)\delta\left(\phi - \phi_i\right) e^{im\phi}d\phi\\
    & = \sum_i \delta\left(R - R_i\right)e^{im\phi_i}.
\label{eq:A_m_det}
\end{split}
\end{equation}
We show the result for the normalised $m=2$ Fourier mode's magnitude in the upper panel of Fig.~\ref{fig:fourier_galaxy_components} for the bulge (orange), disc (blue) and total (black) profiles.
Note that the disc and the total profiles show two peaks: one around $2.5\,$kpc and another one around $6.5\,$kpc.
The first peak arises from the central bar, while the second from the spiral arms. The bulge instead clearly shows only one peak at $\sim 1.5\,$kpc, implying that it also has non-axisymmetric features. 
It is evident from the figure that for $R<4\,$kpc the total profile is influenced by both the bulge and the disc, while at $R>4\,$kpc it follows the disc component only. 

We also show the magnitude and phase profiles for the $A_4$ mode (black dotted lines in Fig.~\ref{fig:fourier_galaxy_components}). The magnitude of this mode is lower than that of the $A_2$ mode for most radii. This indicates that our Galaxy model has a two-fold symmetry with a bar and two spiral arms rather than a four-fold symmetry with a cross and four spiral arms. 
For a bar-like structure -- like the case of our Milky Way simulation -- it is expected that the $A_4$ mode should be about half that of the $A_2$, as is the case in Fig.~\ref{fig:fourier_galaxy_components}.
This is because two axis of the $A_4$ mode's typical cross shape should contribute equally to the total magnitude, but in our case one axis is missing.
In addition, the magnitude of the $A_2$ mode of a cross-like structure would be zero, since the two axes would have opposite phases (as shown in Sec.~\ref{sec:unifom_bar}, a rotation of $90^\circ$ results in a phase shift twice that number for the $A_2$ mode, which results in an opposite phase). Therefore, the dominance of the $A_2$ mode over the $A_4$ mode demonstrates that the galaxy model has a bar and two spiral arms.

The bottom panel of Fig.~\ref{fig:fourier_galaxy_components} shows the phase profile of the Milky Way components. 
We note that unlike for the Fourier magnitude, the phase profile of the total density distribution closely follow that of the disc.
All three lines of the $m=2$ mode display a constant segment at $\pi/3$ that extends up to $\sim2\,$kpc for the bulge and up to $\sim4\,$kpc for the disc. 
The disc profile then decreases down to $-\pi$ at $\sim7\,$kpc, where it suddenly jumps to $\pi$ and then it decreases again. 
This behaviour is characteristic of the barred spiral galaxy as shown in the middle panels of Fig.~\ref{fig:FourierIllustrations}.
In particular, our toy example in Sect.~\ref{sec:barred_spiral} illustrates that the phase of the $A_2$ mode remains constant for the extent of the bar, while the jumps are due to the phase wrapping.

\section{Results} \label{sec:5}

\begin{figure}
   \centering
    \includegraphics[width=1\linewidth]{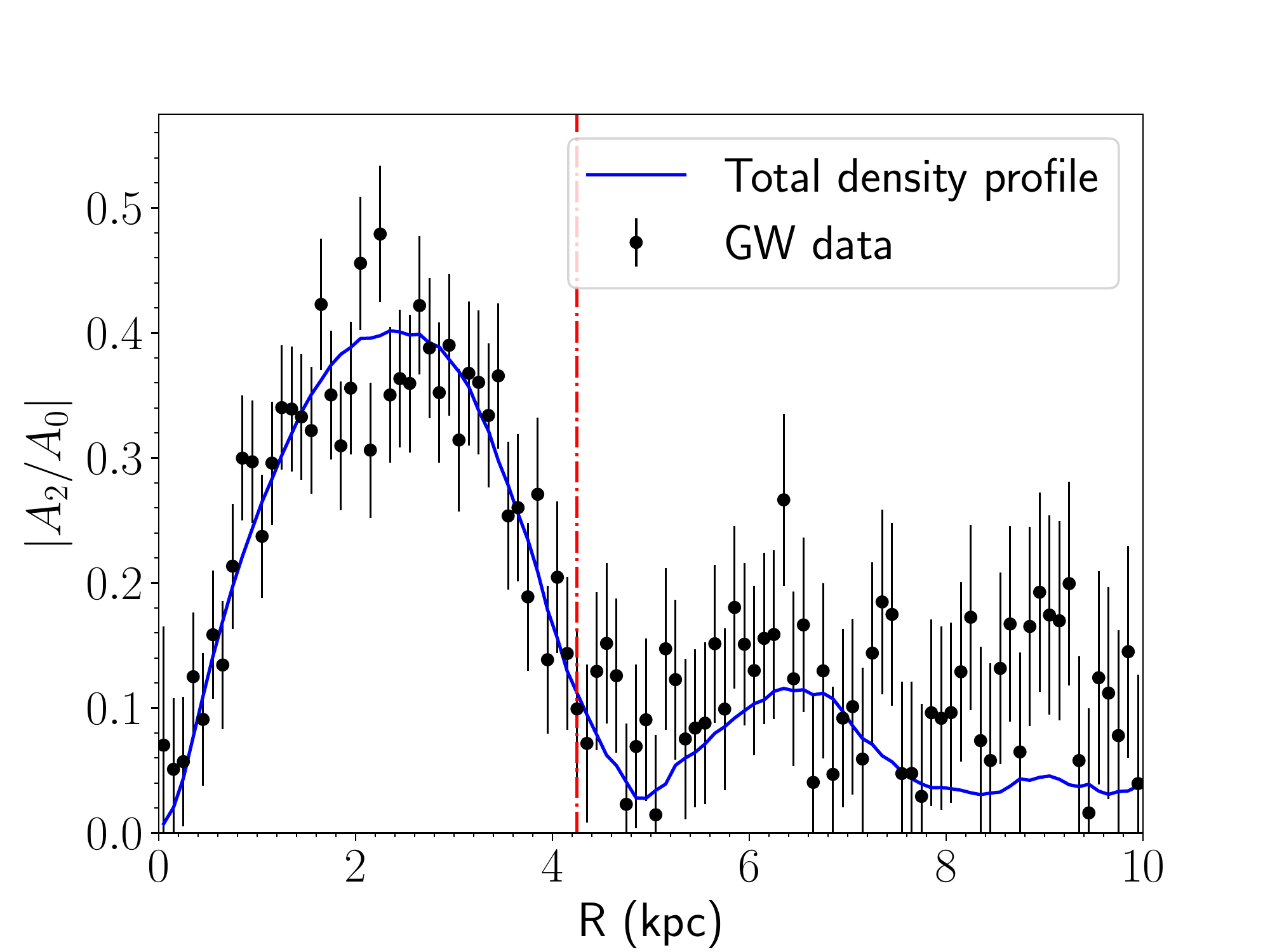}
   \includegraphics[width=1\linewidth]{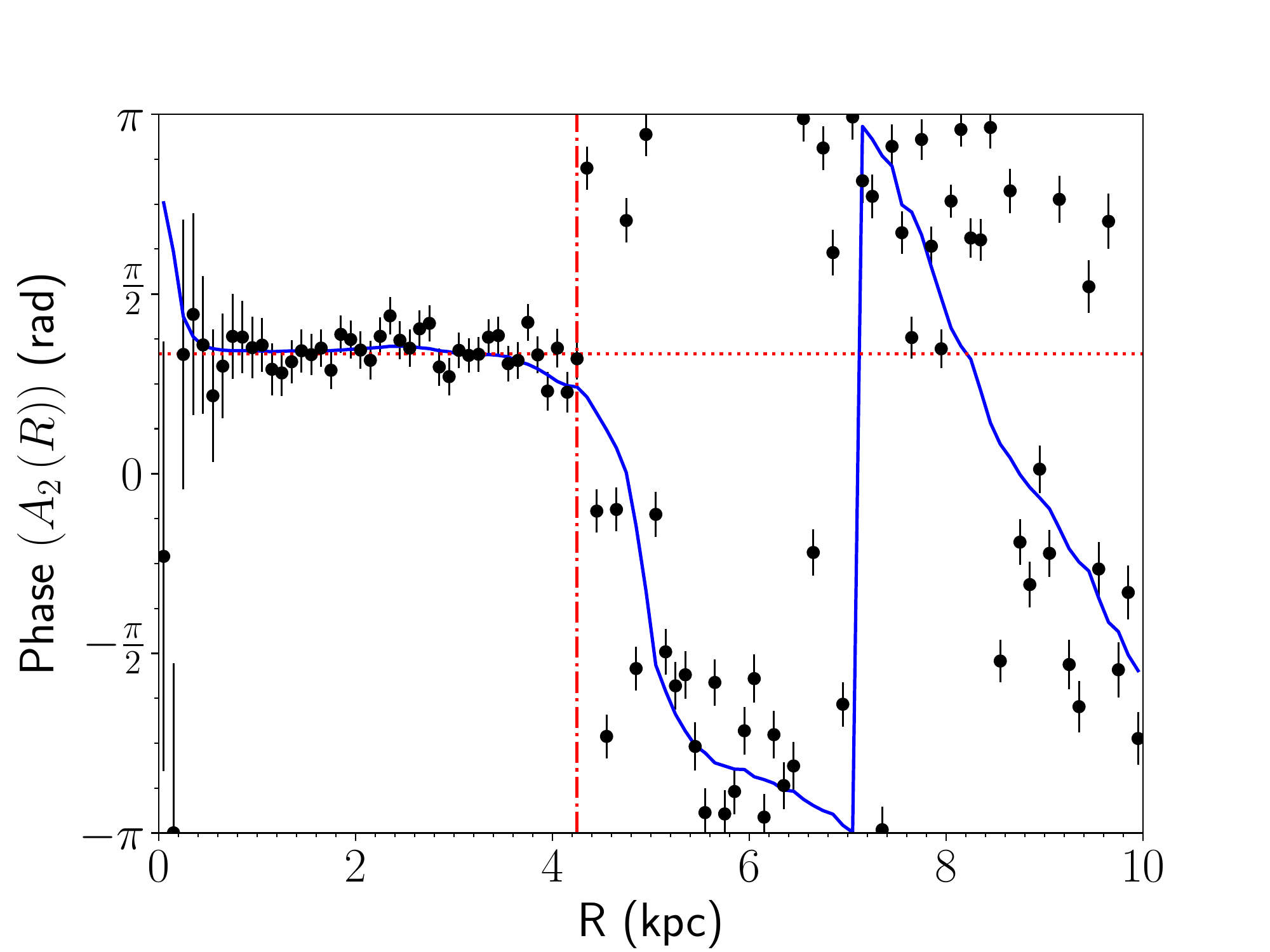}
    \caption{The magnitude and phase of the normalised Fourier mode $|A_2/A_0|$ as a function of the Galactocentric radius of both the total stellar population (blue line, labelled `total density profile')  and the population of detected DWDs, de-biased as described in Section ~\ref{sec:mock_obs} (black points, labelled "GW data"). The error bars represent the standard deviation obtained from $10^{4}$ realisations of the DWD spatial distribution, within the LISA measurement errors.  The vertical, red dash-dotted line denotes a radius of 4.25 kpc, the bar length we recover from the complex phase of the $A_2$ Fourier mode as described in Section \ref{sec:5}. The horizontal, red dotted line in the bottom panel denotes a phase of 60$^\circ$; this is the phase expected for the $m=2$ mode of a bar rotated by 30$^\circ$ (see Section \ref{sec:4}).}
    \label{fig:FC_obs2}
\end{figure}

\begin{figure}
    \centering
    \includegraphics[width=1\linewidth]{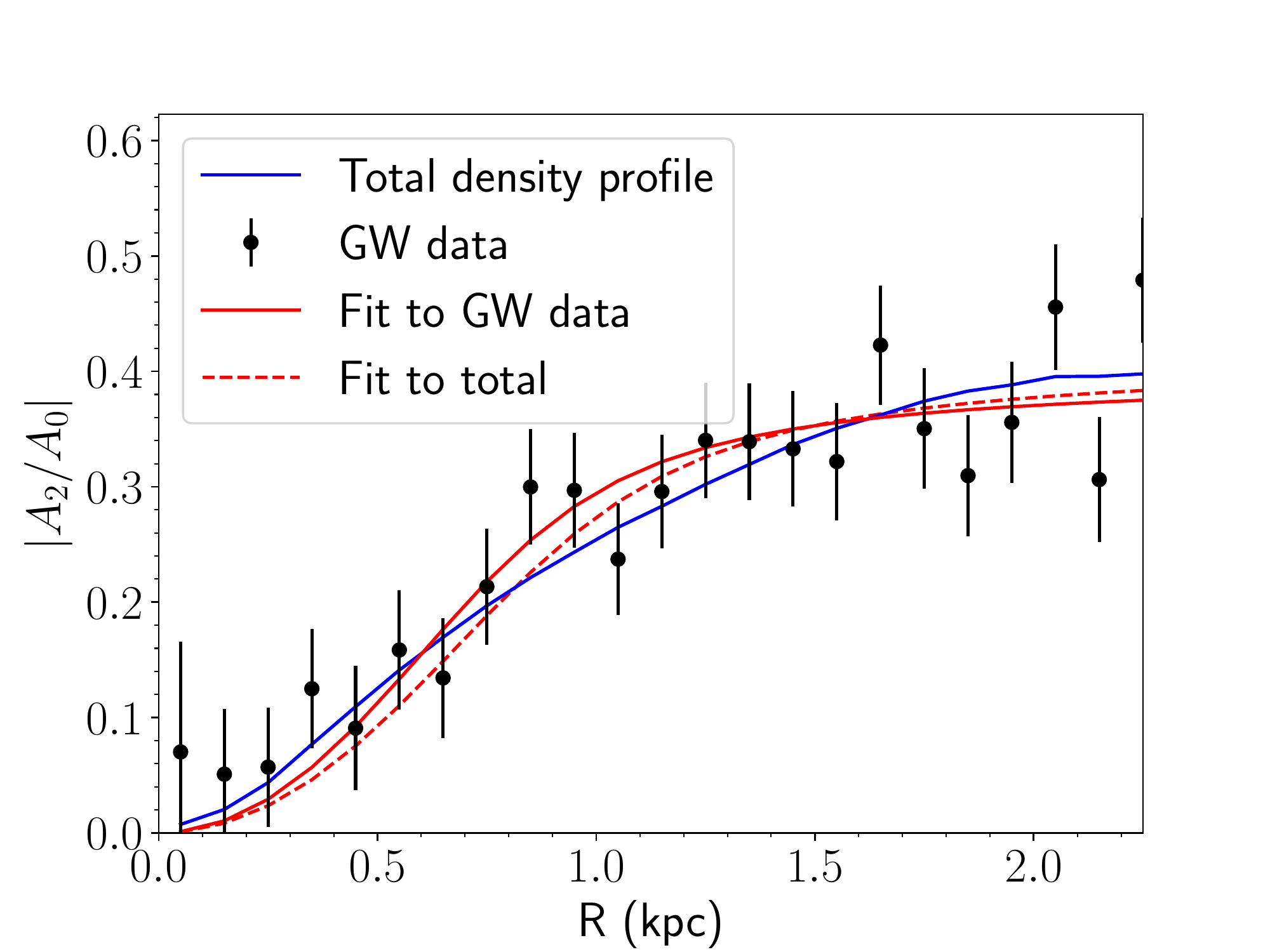}
    \caption{  The magnitude of the normalised Fourier $m=2$ mode of both the full stellar density profile (blue line, labelled total density profile) and the reconstructed population (black points, labelled GW data), as a function of Galactocentric radius. The best fit model to the observations is shown in red solid and the corresponding parameters are bar semi-major scale length $R_{+} = 1.16 \pm 0.05\,$kpc, bar width $R_-=0.31 \pm 0.04\,$kpc, and axis ratio $0.27\pm 0.03$. The best fit to the total density profile is shown in red dashed, and has parameters $R_+ = 1.26 \pm 0.08\,$kpc, $R_-=0.35\pm 0.05\,$kpc, and axis ratio $0.28\pm 0.03$.}
    \label{fig:FC_obs2_fit}
\end{figure}

We calculate the observed Fourier modes (Eq.~\ref{eq:A_m_det}) by radially binning all detections to make a histogram and weighting each detection by its de-biasing weight $w_i$ and a complex phase factor. We choose a bin size $[R,R+dR)$ with $dR = 100$\,pc to obtain the mock GW data (black circles in Fig.~\ref{fig:FC_obs2}).
To quantify error bars, we then repeat the procedure for $10^4$ realisations of the binary positions in the Galaxy. Specifically, each realisation is obtained by randomly drawing distances and sky positions from Gaussian distributions centred on the measured values, with standard deviation equal to the measurement errors provided by the MLDC pipeline.  This procedure allows us to assign statistical uncertainties to a given realisation, calculated as the standard deviation of all realisations (error bars in Fig.~\ref{fig:FC_obs2}).

Figure~\ref{fig:FC_obs2} illustrates the obtained normalised $m=2$ Fourier mode's magnitude (top panel) and phase (bottom panel) as a function of the Galactocentric cylindrical radius. 
For comparison we over-plot the Fourier transform of the total stellar distribution in the Galaxy (cf. Sect.~\ref{fourier mw}) using  the blue solid line.
In the top panel the mock data follow the total density profile well up to roughly $7\,$kpc, describing well the first peak.
Beyond $7\,$kpc the data become too noisy to identify over-densities.
In the bottom panel the mock data closely follow the phase of the total density profile (blue line) up to $\sim5\,$kpc and become more scattered at greater radii. As can be seen in Fig.~\ref{fig:GalMaps1}, this is due to the low density contrast between DWDs in the spiral arms and the background disc.
From a visual inspection of Fig.~\ref{fig:FC_obs2} we can conclude that by using the reconstructed DWD density maps from LISA observations one can clearly identify only the bar. The spiral arms' identification and characterisation is more challenging than for the bar, whether we exploit the $m=2$ Fourier magnitude or the phase.

\subsection{Geometric parameters of the bar}

To derive the bar's parameters, we fit the mock data with the Fourier transform of the 2D-Gaussian profile defined in Eq.~\eqref{sec:gauss_bar}.
The fit is performed from the origin up to and including the maximum magnitude of $A_2/A_0$ that we arbitrarily identify as the maximum data point in the top panel of Fig.~\ref{fig:FC_obs2}. 
Specifically, we fit for $R_{\rm x}$ (the scale length of the Gaussian bar in the $X$-direction) and $k$ (parameter that encodes the slope of the first peak in the magnitude profile).
We note however that the fit of $R_{\rm x}$ is influenced by our choice of the data point representing the maximum of the $|A_2/A_0|$ profile, while our result for $k$ is relatively insensitive to the this position as long as all data points on the rising edge are included.
Fig.~\ref{fig:FC_obs2_fit} shows the inner ~2.5 kpc of the top panel of Fig.~\ref{fig:FC_obs2}, with the best fit Fourier-transformed Gaussian for comparison. 

We find $k$ to be $> 1$, and therefore derive from Eqs.~\eqref{eq:rb}-\eqref{eq:h} the `observed' axis ratio as ${1}/{\sqrt{k}} = 0.27 \pm 0.03$.
We derive the `true' axis ratio by fitting the Fourier magnitude profile of the total stellar mass distribution that yields $0.28\pm0.03$.
The `observed' and `true' values of the bar axis ratio agree within $1\sigma$. 
We recover
`observed' scale lengths $R_{+} = 1.16 \pm 0.05\,$kpc, and $R_-=0.31 \pm 0.04\,$kpc. We compare these values to the fit to the Fourier magnitude profile of the stellar mass distribution that yields $R_{+} = 1.26 \pm 0.08\,$kpc, and $R_-=0.35\pm0.05\,$kpc.  
Thus, the observed value for the bar semi-minor scale length differs by about $1\sigma$ with respect to the true value, and the bar semi-major scale length $R_{+}$ differs by about $2\sigma$.

As discussed in Sec.~\ref{sec:4}, the bar length and orientation can be independently recovered from the phase of the $m=2$ mode. This is possible using the fact that the phase stays constant at $\Phi\left( A_2\right)=2\phi_0$ for the extent of the bar and deviates from this value  beyond it (see Fig.~\ref{fig:FC_obs2}). Additionally, because of the relative scarcity of detected sources outside of the bar, the data follow the model much better in the region dominated by the bar.
To perform these measurements we apply the following criterion. Between 0.5\,kpc (to exclude the bulge) and $R$ we fit a horizontal line $\Phi \left(A_2\right) = {\rm const}$ and compute the $\chi^2$ value of the fit up to that given point. In this way, we obtain a function $\chi^2\left(R\right)$ that quantifies the goodness of fit to a constant value, for data up to $R$.
Using the statistical distribution of $\chi^2$ we can then compute the cumulative distribution function (CDF) for each fit, that can be interpreted as the probability that a randomly generated set of data fits the model better. We compute 1 - CDF, i.e. the probability of generating a random set of data from the model that fits the model worse. We then have a function that at every $R$ quantifies how well a constant can fit points up to that $R$. The bar length is then defined as the $R$ up to which the fit is good. We take a lower limit of 0.05\footnote{i.e. the probability that a set of data randomly generated from the model fits the model worse than this data is 5\,per cent} for an acceptable fit. We find that for $R \leq 4.25$\,kpc, our statistic is always greater than 0.6, while for $R > 4.25$\,kpc, it is always smaller than $10^{-6}$. The boundary thus appears well-defined, and relatively insensitive to our choice of the threshold probability value. 
In this way we recover a bar length of $R_{\rm b}=4.25\,$kpc. We retrieve the viewing angle by taking the best fit to the data up to the bar length, which gives us a value of $\phi_0 = 30.70\pm0.83^\circ$, in agreement with the true viewing angle of 30$^\circ$.

The vertical red dash-dotted lines in both panels of Fig.~\ref{fig:FC_obs2} indicates this bar length of 4.25 kpc. Visual inspection of the bottom panel shows that this line coincides with a `knee' in the phase profile of the $m=2$ Fourier mode, which we argue is indeed the point where the bar transitions into the spiral arms. 

We again stress that this method locates the transition between regions of good and bad reconstruction, rather than the turnoff from constant phase. Both of these are characteristic of the transition from the bar to the disc (although it is assumed that LISA does not reconstruct the disc well). However, the first method requires more careful statistical treatment as the $\chi^2$ statistics assumed are not sufficient. The position information of the data points is not used when computing the $\chi^2$ value, while the feature to be identified is a deviation in a very specific region, in a very specific sequence.

 We can reconstruct the dimensions of the bar by combining the results presented above. The axis ratio derived from the scale lengths of the main axes (obtained using the magnitude of the Fourier $m=2$ mode) can be combined with the physical length (obtained using the complex phase of the Fourier $m=2$ mode) to determine the physical width of the bar. In analogy to the uniform density bar in Sec.~\ref{sec:unifom_bar} we define the full bar width as $H=2R_{\rm b}/\sqrt{k}$. We find $H = 2.28 \pm 0.25$\, kpc.

\begin{figure*}
   \centering
   \includegraphics[width=0.8\linewidth]{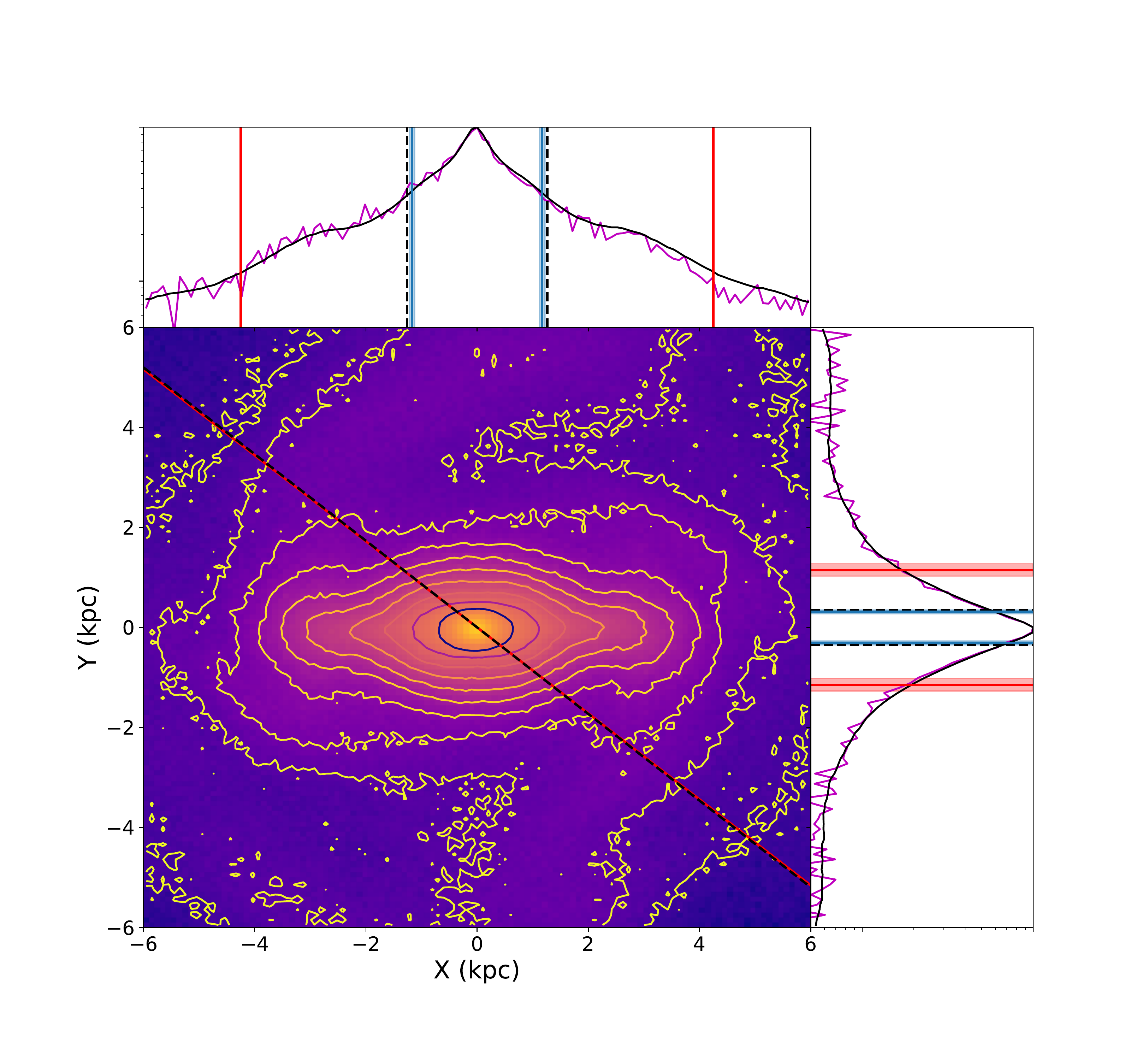}
    \caption{The total stellar mass density map in the bar's reference systems with contours of constant density. The top and right panels show the marginalised density profiles of the full stellar density distribution (black) and the observed DWD density distribution (magenta) scaled to the same maximum. Also shown are the bar (scale) lengths recovered from the magnitude (in blue, with $1\sigma$ confidence intervals), and from the phase (in red, using the axis ratio from the magnitude to find the bar width; error bars are $1\sigma$ confidence intervals due to errors in the axis ratio). The black dashed lines denote the reference values for the bar scale lengths as found from fitting to the magnitude profile of the full stellar distribution. Finally, the red solid and black dashed lines across the density map denote the viewing angle recovered from the phase and the reference value. Note that the Sun is located off to the right of this plot, on the black dashed line indicating our line-of-sight. 
    }
    \label{fig:Galaxy_BarCircle_V2}
\end{figure*}

In Fig.~\ref{fig:Galaxy_BarCircle_V2} we summarise our results by comparing our recovered bar parameters to the total density distribution in the Galactic plane in the reference frame aligned with the bar.        
The density contours (spaced by 0.17 dex) in the central panel give us an idea of the extent of the bar: a number of contours assume a close boxy shape in the centre extending between about -4.5\, kpc and 4.5\, kpc in the $X$ direction and between about -2.5 and 2.5\, kpc in the $Y$ direction.
In the adjoined panels we show the marginalised density profiles along $X$ and $Y$ directions and we indicate our estimates with vertical lines.
We note that red lines in the top panel align well with the last `boxy' contour in the middle panel, while the red lines on in the right panel are narrower than the density contours. 

\subsection{Dependence on viewing angle} \label{sec:viewing_angle}

In order to investigate the sensitivity of our results to different viewing angles  we perform a set of 13 simulations for a number of vantage points from 0 to $180^{\circ}$. In all simulations we place the virtual LISA detector at $8.1\,$kpc from the Galactic Centre and rotate it around a half-circle in steps of 15$^{\circ}$. 
For each viewing angle, the de-biasing function has been recomputed using Eq.~\eqref{eq:bias}. We report different statistics for the obtained the power law indices and scaling constants in Table~\ref{tab:debiasing}. We note that their standard deviations are of the order of a few per cent and their means and medians agree to about 10\,per cent of a standard deviation, implying that these are robust results.  These values can be potentially used to construct a de-biasing function for actual observations of DWDs with LISA. However, the degree to which our de-biasing function approximates the true one depends on the quality of the LISA noise model and on the DWD population model, and warrants a separate more careful investigation.

\begin{figure}
    \centering
    \includegraphics[width=1\linewidth]{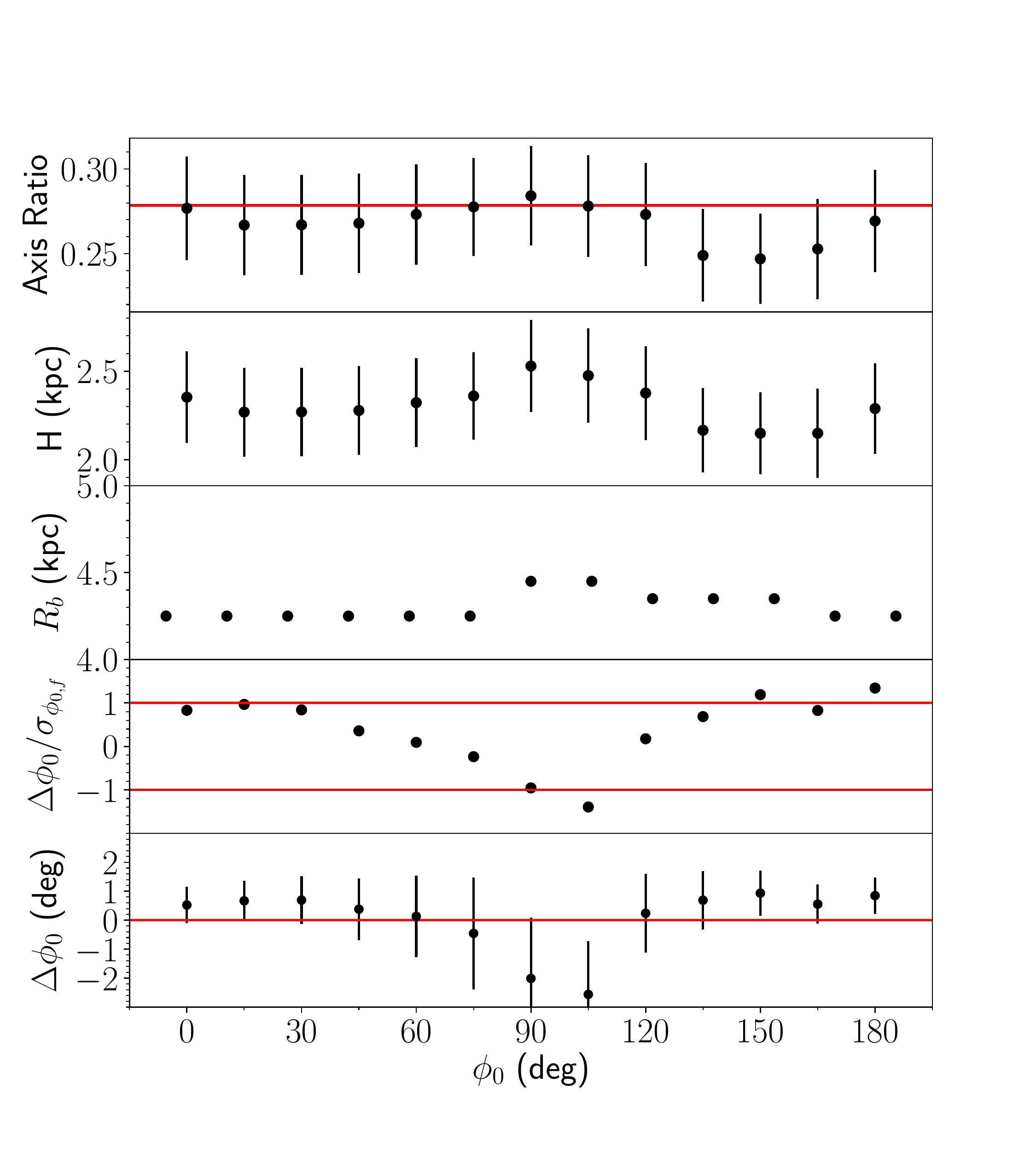} 
    \caption{Estimated bar's structural properties as a function of the viewing angle. From top to bottom we show: axis ratio, bar's physical width $H$ and length $R_{\rm b}$, relative viewing angle error and absolute viewing angle error (with $\Delta\phi_0$ being the difference between the fitted and true viewing angle). Error bars indicate $1\sigma$ fitting errors. Red horizontal lines illustrate reference values where available. In the top panel, it is the value derived from fitting the Gaussian model to the total stellar distribution. In the second panel from the bottom, it indicates an error of $1\sigma$. In the bottom panel, it is the line $\Delta\phi_0=0$. }
    \label{fig:AngleVar_Rb_H}
\end{figure}

We illustrate the obtained results for the bar's structural parameters for each simulation (black circles) in Fig.~\ref{fig:AngleVar_Rb_H}. 
Where possible, we use a horizontal red line to illustrate a reference value.
We note that axis ratios (derived from the magnitude of the $m=2$ Fourier profile) are generally consistent with the true value and with each other within $1\sigma$. The bar widths (derived from both the magnitude and phase of the $m=2$ Fourier mode) are generally consistent with each other within $1\sigma$.
The fitted viewing angles are for the most part consistent with their true values, with errors on the order of $1^\circ$.

\begin{table}

    \caption{Statistical properties of the parameters of the de-biasing function, for the 13 viewing angles (0$^\circ$ up to and including 180$^\circ$, in steps of 15$^\circ$).
    }
    \label{tab:debiasing}
    \begin{center}
    \begin{tabular}{ccc}
        \hline
        Parameter & Scaling & Power law index \\
        \hline
        \hline
        Range & [0.978e2, 1.092e2] & [0.907, 0.967] \\
        Mean & 1.027e2 & 0.938 \\
        Standard Deviation & 0.03e2 & 0.016 \\
        Median & 1.022e2 & 0.940 \\
        \hline
    \end{tabular}
\end{center}
\end{table}

\section{Discussion and conclusions} \label{sec:6}

It is well established that the Milky Way presents a complex baryonic structure including bulge, a thin and a thick disc and a stellar halo \cite[see][for a review]{bla16}.
In particular, the structure of the inner Galaxy is not well known due to heavy extinction and stellar crowding. The best available structural information on the stellar population of the inner Galaxy comes from large samples of Red Clump Giant stars visible in near-infrared band, for which individual distances can be determined \citep{min10a,weg13,simion_parametric_2017}. 
Numerical simulations show that bulges represent the inner part of a longer, planar bar structure that formed through buckling out of the galaxy plane and/or from orbits in vertical resonance \cite[e.g.,][]{ath05}. Therefore our Galaxy is also expected to have a thin bar component extending well outside the bulge. However, finding the Galactic planar bar and characterising its properties has proven difficult, again because of intervening dust extinction and the superposition with the star-forming disc at low latitudes toward the inner Galaxy.

In this paper, we exploit the fact that GW observations do not suffer from these effects and that GW strength decreases slower with distance than electromagnetic waves, allowing us to reach the far side of the Galaxy. In particular, we propose and develop for the first time a methodology for investigating the inner structure of the Milky Way using future LISA observations. We combine binary population synthesis techniques with a high resolution Milky Way simulation for a realistic modelling of the Galactic structure \citep{don19}.

Our results reveal that the shape of the bar is already apparent by simply plotting the 3D positions of LISA detections (unlike with optical observations \citet{and19}), while the spiral arms remain elusive.
To recover the bar's parameters, we analyse the density distribution in Fourier space of LISA detections. We find that the $m=2$ Fourier mode is the strongest, implying the presence of a bar plus a pair of spiral arms structure. We note that our choice to analyse the $m=2$ mode is also motivated by the space distribution of the LISA detections (cf. Fig.~\ref{fig:GalMaps1}) and is focused on the central region of the Galaxy containing the bar; our analysis does not {\it a priori}  exclude a multi-arm spiral structure at larger Galactocentric radii. Moreover, numerical simulations show that by the time the bar grows in strength and size the only remaining prominent spirals are the ones located at the edge of the bar with subdominant arms in the outer disc \citep{don15} - supporting our choice to consider the $m=2$ Fourier mode.

By fitting a 2D Gaussian density function to the Fourier transform of the bar, we recover a semi-major scale length $R_{+}=1.16 \pm 0.05\,$kpc, a semi-minor scale length $R_-=0.31\pm0.04\,$kpc and an axis ratio $0.27\pm0.03$. The result for the bar axis ratio is consistent within $1\sigma$ with the true values, obtained by performing the same analysis on the total stellar population. The semi-minor and semi-major scale lengths, instead, are underestimated by $\sim 1 \sigma$ and $\sim 2 \sigma$ respectively, although the best-fit and true values differ by just $\sim 10$ per cent.
The axis ratios obtained assuming different viewing angles (cf. Fig.~\ref{fig:AngleVar_Rb_H}) are generally consistent with each other within $1\sigma$ and with the true value. Although Fig.~\ref{fig:FC_obs2_fit} shows that the Gaussian model does not provide an exact match to the density profile of the bar, 
it still allows us to recover the axis ratio in a robust way as shown in Sect.~\ref{sec:viewing_angle}. 
By fitting the phase of the $m=2$ mode as a function of $R$, we recover a viewing angle $\phi_0 = 30.70 \pm 0.83\,^{\circ}$, consistent with the true angle within $1\sigma$. Viewing angle estimates inferred from other vantage points (Fig.~\ref{fig:AngleVar_Rb_H}) are also generally consistent with the true value within $1\sigma$; the fitting error is of the order of $1^\circ$. Likewise, the bar lengths recovered with this method for different viewing angles do not vary by more than 0.2 kpc, or 5 per cent. By combining the axis ratio recovered from the magnitude and the bar length recovered from the phase we recover a bar width that visually fits well to the model. This lends further credence to the axis ratio derived with our method.

Let's now compare our results with the state-of-the-art results derived using electromagnetic tracers. Electromagnetic studies of the geometric properties of the Galactic bar have used optical and infrared stellar tracers. Recent examples include infrared observations of stars using the 2MASS survey \citep{robin_stellar_2012}, optical observations of Red Clump stars \citep{cao_new_2013} and RR Lyrae stars \citep{pietrukowicz_deciphering_2015} from the OGLE survey, and optical observations of Red Clump stars with the VVV survey \citep{weg13,simion_parametric_2017}. The geometric parameters of the bar obtained by these five studies vary between $10^\circ - 45^\circ$ and $0.25 - 0.6$ in viewing angle and in-plane axis ratio respectively  \citep[][figure ~17]{simion_parametric_2017}.  
Error estimates are reported to be of the order of a few degrees for the viewing angle \citep{robin_stellar_2012,pietrukowicz_deciphering_2015,weg13}; around tens of per cent for scale lengths \citep{robin_stellar_2012}; and around \textasciitilde 5\,per cent for the in-plane axis ratio \citep{pietrukowicz_deciphering_2015}.
Instead from GWs measurements, one can recover the bar in-plane axis-ratio, physical length and viewing angle accurately (within $1\sigma$ from the true values, where available) and with a precision of $< 10\,$ per cent for the two former quantities, and 1$^\circ$ for the latter one. Therefore, GWs promise to be a competitive observational window for the central region of the Galaxy.

Nevertheless, the analysis in this paper can be refined in a few ways. We have assumed a simple Galactic plane-projected, bi-axial Gaussian density profile, whereas the Milky Way's bar is more peanut-shaped and has a vertical structure that we have ignored in this work. The fact that this model is insufficient is evident from the fact that the red dashed line in Fig.~\ref{fig:FC_obs2_fit}, representing the model fit to the total density distribution, does not follow the total density distribution very well, and that the bar scale lengths are underestimated. Additionally, we didn't attempt to simultaneously recover the scale height of the bulge and disc, although possible as shown by \citet{ada12} and \citet{kor19}, albeit with simple analytical density models. Finally, our method for determining the bar length and viewing angle is based on a very simple statistical method and works because of the density contrast between bar and disc. A more detailed and universal analysis is certainly possible. 

Another improvement pertains to our choice for the star formation rate. While we assumed that the bar follows the same star formation as the disc, one should instead self-consistently simulate the star formation history of the bar itself. The different star formation histories of the disc and bulge, and bar can potentially give rise to two distinct populations. Finally, the future is currently offering at least two more space-based interferometers beside LISA: the DECI-hertz inteferometer Gravitational wave Observatory (DECIGO, \citealt{Sato+17}) and TianQin (\citealt{Luo+16}). An analysis that would consider the combined measurements would likely result in a more effective and precise tomography of the Galaxy.

Presented results depend on removing the detection bias arising from the dependence of the SNR on the distance. Our de-biasing method (Sect.~\ref{sec:bias}) relies on the assumption of a homogeneous DWD population throughout the Milky Way, with a frequency distribution only determined by GW emission. In reality, different regions of the Milky Way have undergone different star formation histories, locally modifying the DWD binary frequency distribution \citep[e.g.][]{yu13,lam19}.
Including this would require a substantial extension of our modelling and it is left for a future follow-up. Here, however, we can test alternative ways to determine the bar's parameters, that do not rely on the de-biasing procedure.
Specifically, we use the raw sample of all observed DWDs -- those that factually pass the SNR threshold -- and a subset of it, that only includes sources emitting at frequencies $\geq 3$\,mHz. 
Around this frequency, LISA is most sensitive and the required signal for detection can be $\sim10\times$ lower than at $\sim 10^{-4}$\,Hz, allowing detections at $10\times$ larger distances. This high-frequency DWD population therefore suffers far less from distance biases at Galactic scales considered here \citep{kor20,roe20}.
We find that the high frequency sample contains about a third of the LISA detections.

\begin{figure}
    \centering
    \includegraphics[width=1\linewidth]{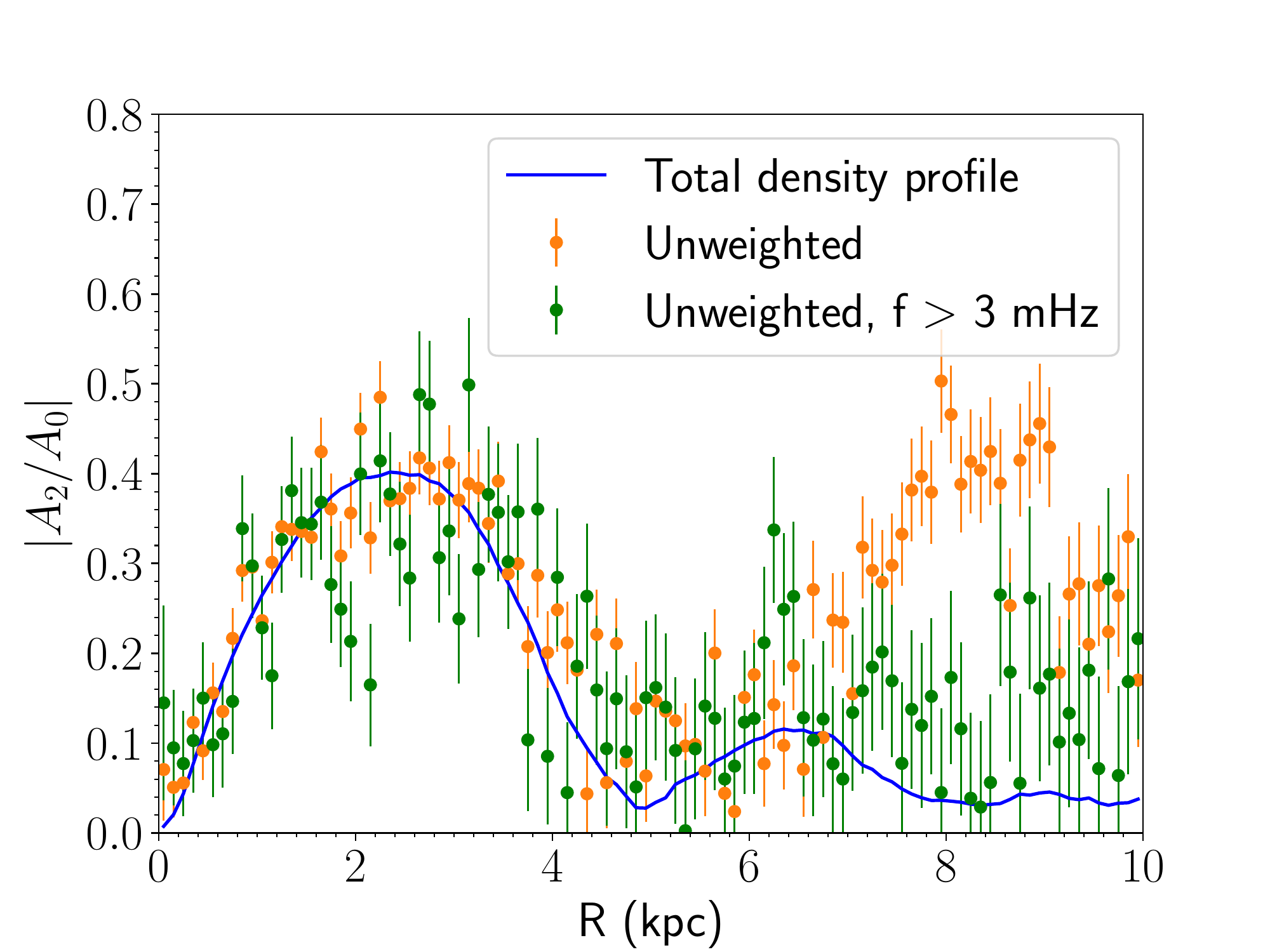}
    \includegraphics[width=1\linewidth]{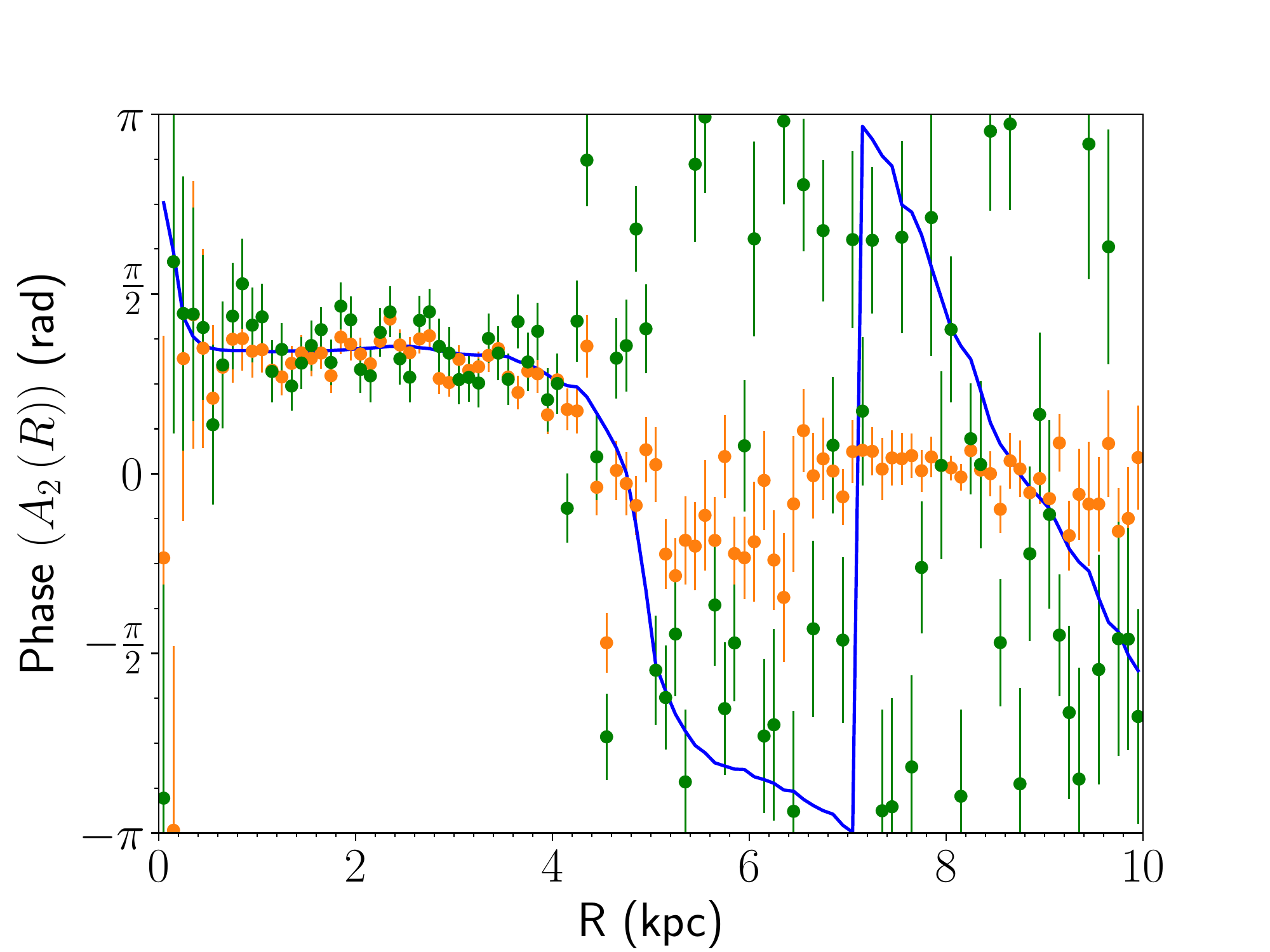}
    \caption{The magnitude and complex phase of the normalised $m=2$ Fourier mode of the full observed DWD population (blue solid line), without de-biasing weights (orange circles), and all DWDs with $f>3\,$mHz (green cycles) without de-biasing weights. The error bars represent the standard deviations obtained from $10^4$ realizations of the DWD spatial distribution, within LISA measurement errors.}
    \label{fig:FC_obs_2_discus}
\end{figure}

The Fourier mode $A_2$ magnitude and phase profiles for these two samples are shown in Fig.~\ref{fig:FC_obs_2_discus}. 
The magnitude profile of the full observed sample (unweighted, orange circles) follows well the peak due to the presence of the bar, and shows a secondary peak around $8$ kpc. This corresponds to an over density of observed systems around the Sun, that is not mirrored anywhere else at the same distance (cf. Fig.~\ref{fig:GalMaps1} top panel).
In these conditions the Fourier transform will show a non-zero amplitude at the -- non symmetric -- overdensity location. This is visible also in the behaviour of the phase (bottom panel of Fig.~\ref{fig:FC_obs_2_discus}): beyond 6.5 kpc, the phase is about $\sim 0$ that implies a structure on the line of sight to the Galactic Centre. 
In comparison, the magnitude and phase profiles of high-frequency sources (green circles in Fig.~\ref{fig:FC_obs_2_discus}) are similar to those of the de-biased sample (Fig.~\ref{fig:FC_obs2}), although significantly noisier.  

Following the same procedure as described in Sect.~\ref{sec:5},
by fitting the Fourier magnitude profile from the unweighted data we recover: an axis ratio $0.27\pm 0.01$, $R_{+} = 0.71\pm 0.0002$, $R_-=0.19\pm 0.01$; from the frequency limited sample we obtain: an axis ratio $0.25\pm 0.03$, $R_{+} =1.24\pm 0.02$, $R_-=0.31\pm 0.04$. 
Of these, both axis ratio's agree with the reference value, and both the bar semi-major and semi-minor axes from the full dataset disagree with the reference value by many standard deviations. The bar semi-major and semi-minor scale lengths from the high frequency subset agree within $1\sigma$. 

Similarly, from the phase profile of the unweighted data we recover $R_{\rm b} = 4.35$ kpc and $\phi_0 = 28.09\pm 0.72^\circ$. From the unweighted, high frequency data we recover $R_{\rm b} = 4.05$ kpc and $\phi_0 = 31.00\pm 1.13^\circ$. In both cases, the error in the bar length is of the same order as the dispersion among the different viewing angles. The viewing angle derived from the unweighted data differs from the reference value by about $3\sigma$, whereas the one derived from the high frequency data agrees within one standard error (albeit with a larger standard error than for the weighted data). 

We therefore conclude that the best alternative sample to the de-biased one is that of observed high-frequency DWDs, that can be exploited to compare and validate results. 
In particular, the high-frequency sample yields precise and accurate measurement of the bar's axis ratio. However, to accurately recover the bar length and viewing angle from the phase profile, the de-biasing process is required.

Despite these limitations, our analysis represents a first proof of concept of the feasibility to recover and study the inner parts of our Galaxy using GWs, and we anticipate that a synergistic study with more traditional electromagnetic tracers will be a new and very effective Galactic investigation strategy in the 2030s. 

\section*{Acknowledgements}

\addcontentsline{toc}{section}{Acknowledgements}
We thank Silvia Toonen for providing us the model population for DWD, Nilanjan Banik and Tyson Littenberg for useful discussions.\\ This work was supported by NWO WARP Program, grant NWO 648.003004 APP-GW.
VK acknowledge support from the Netherlands Research Council NWO (Rubicon 019.183EN.015). E.D.O acknowledges support from the Vilas Reaserach Fellowship provided by University of Wisconsin.\\
This research made use of {\sc NumPy, SciPy} and {\sc PyGaia} python packages, matplotlib python library and the publicly available pipeline for analysing LISA data developed by Tyson Littenberg and Neil Cornish available at \url{github.com/tlittenberg/ldasoft}. 

\addcontentsline{toc}{section}{Acknowledgements}

\section*{Data availability}

\addcontentsline{toc}{section}{Data availability}
The data underlying this article is publicly available at \url{https://figshare.com/articles/dataset/DWD_FullPopulation_Wilhelm_Korol_Rossi_DOnghia_2020/13168464}.

\addcontentsline{toc}{section}{Data availability}





\bibliographystyle{mnras}
\bibliography{biblio.bib}




\bsp	
\label{lastpage}
\end{document}